\documentclass[12pt]{article}

\usepackage{scicite}
\usepackage{graphicx}
\usepackage{times}
\usepackage{amsmath}
\usepackage{xcolor}
\usepackage{soul}
\usepackage{float}
\usepackage{hyperref}
\usepackage{amssymb}
\usepackage{xcolor}

\usepackage{draftwatermark} 

\newenvironment{sciabstract}{%
\begin{quote} \bf}
{\end{quote}}

\title{ Microstructure and Stress Mapping in 3D at Industrially Relevant Degrees of Plastic Deformation }

\author
{Axel Henningsson,$^{1\ast}$ Mustafacan Kutsal,$^{2,3}$ Jonathan P. Wright,$^{3}$ \\ Wolfgang Ludwig,$^{3,4}$ Henning Osholm Sørensen,$^{5}$ Stephen A. Hall,$^{1}$ \\ Grethe Winther$^{6}$ and Henning F. Poulsen$^{2\ast}$ \\
\\
\normalsize{$^{1}$Division of Solid Mechanics, Lund University, Ole Römers Väg 1, Lund, Sweden,}\\
\normalsize{$^{2}$Department of Physics, Technical University of Denmark, Kongens Lyngby, Denmark}\\
\normalsize{$^{3}$European Synchrotron Radiation Facility, 71 Avenue des Martyrs,}\\
\normalsize{CS40220, 38043 Grenoble Cedex 9, France} \\
\normalsize{$^{4}$MATEIS, Universite de Lyon, INSA Lyon, CNRS UMR5510,}\\
\normalsize{69621 Villeurbanne, France} \\
\normalsize{$^{5}$Xnovo Technology ApS, Galoche Allé 15, 1st floor, 4600 Køge, Denmark} \\
\normalsize{$^{6}$Department of Civil and Mechanical Engineering,}\\
\normalsize{Technical University of Denmark, Kongens Lyngby, Denmark}\\
\normalsize{$^\ast$To whom correspondence should be addressed;} \\ \normalsize{E-mail:  axel.henningsson@solid.lth.se \& hfpo@dtu.dk}
}

\date{}

\begin{document}


\baselineskip24pt


\maketitle 

\SetWatermarkText{Preprint} 
\SetWatermarkScale{1} 

\begin{sciabstract}
    \textcolor{red}{This is a preprint of a paper that is currently in submission for publication. For citation, please refer to the peer-reviewed published work whenever possible.}
\end{sciabstract}

\begin{sciabstract}
    Strength, ductility, and failure properties of metals are tailored by plastic deformation routes. Predicting these properties requires modeling of the structural dynamics and stress evolution taking place on several length scales. Progress has been hampered by a lack of representative 3D experimental data at industrially relevant degrees of deformation. We present an X-ray imaging based 3D mapping of an aluminum polycrystal deformed to the ultimate tensile strength (32\% elongation). The extensive dataset reveals significant intra-grain stress variations (36 MPa) up to at least half of the inter-grain variations (76 MPa), which are dominated by grain orientation effects. Local intra-grain stress concentrations are candidates for damage nucleation. Such data are important for models of structure-property relations and damage.
\end{sciabstract}

The utilization of polycrystalline metals and alloys is widespread across various industries such as modern electronics, automotive, construction, aerospace, and energy. Notably, through plastic deformation of a polycrystalline specimen, a permanent alteration of its shape with a spectrum of macroscopic properties can be effectively controlled. These properties encompass mechanical strength, hardness, and ductility, alongside electrical and thermal conductivity, as well as magnetic coercivity and permeability.
To elucidate the mechanisms behind the evolution of these macroscopic properties with plasticity, an examination of the microscopic interplay both between and within the single crystal grains constituting the polycrystalline aggregate is imperative. This interplay is characterized by the local crystal orientation and stress state. Understanding this interplay across various length scales is pivotal for addressing fundamental scientific inquiries related to damage, creep, deformation twinning, and notably, the correlation between strength and processing.
Consequently, there exists a significant demand for a quantitative mapping of the 3D microstructural and stress development of the polycrystal in response to external forces. Moreover, from an engineering safety standpoint, mapping plastic deformations at both inter- and intra-grain length scales is critical as they precede component failure.

Electron microscopy is widely employed to map the microstructure of metals. Transmission electron microscopy (TEM) enables nanometer-resolution 3D mapping of crystal orientation and defect structures within thin films \cite{Liu2011, Harrison2022, He2023} and micro-pillars \cite{Uchic2004}. Concurrently, electron backscatter diffraction (EBSD) combined with serial sectioning facilitates the mapping of local orientation within extensive volumes \cite{Uchic2006, Echlin2011}. 
However, these techniques fall short in tracking the dynamic evolution of the microstructure \emph{in situ} in a manner representative of bulk behavior. Similarly, stress mapping \cite{Wilkinson2006, Dingley2018} faces challenges due to geometrical constraints arising from the limited penetration power of electrons. In contrast, X-ray diffraction based imaging methods such as 3D X-ray Diffraction \cite{Poulsen2001}, Diffraction Contrast Tomography \cite{King2008, Ludwig2009} and lab-DCT \cite{McDonald2015} excel in producing voxelated maps of the lattice orientation in deformed metals \cite{Li2012, Evan2016} in mm-sized samples \cite{Margulies2001} comprising up to tens of thousands of grains \cite{Pokharel2015, Rachel2022}. 
Nevertheless, the complementary mapping of the intra-grain stress \cite{Jakobsen2006} has proven elusive, due to the increasing overlap of diffraction patterns from different parts of the sample as deformation progresses. Such maps are critically important in guiding and validating the modeling of plasticity and failure.

Introducing a monochromatic beam raster scanning modality, scanning 3DXRD (S3DXRD), reduces these limitations. By adapting reconstruction algorithms from tomography, complimentary mappings of intra-grain residual orientation and stress have been made \cite{Bonnin2014, Hektor2019, Henningsson2020}. This advancement has been further augmented through the use of conical slits \cite{Nielsen2000}, which physically reduce the diffracting gauge volume \cite{Hayashi2019,Hayashi2023SpiralSlits}. As examples of results, residual stresses have been mapped in a plastically deformed titanium sample at \(7\)\% tensile deformation \cite{Li2023} and in a low carbon steel at \(5.1\)\% tensile deformation \cite{Hayashi2019}. However, to our knowledge, no 3D bulk reconstructions of intra-grain stress have been achieved at higher levels of deformation.  In contrast, industrially relevant processes for metals, such as rolling, extrusion, drawing and forging typically require medium to high levels of deformation (10-90 \%). Likewise, in simple tensile testing, most ductile metals may be elongated 20-40\% before the onset of failure.

In order to push the deformation limit of the nondestructive 3D stress mapping, here, we present a novel reconstruction algorithm designed to address the challenges posed by diffraction spot overlap. Leveraging spatial correlation within the orientation field, this algorithm effectively mitigates noise. We demonstrate its use on a coarse-grained aluminum alloy subject to \(>30\)\% tensile deformation in an S3DRXD set-up at the European Synchrotron Radiation Facility (ESRF). The sample is deformed to its ultimate tensile strength (UTS) where plastic deformation localizes and damage is initiated. 
We mapped the orientation and stress tensor in $\approx 250,000$ voxels in a volume of $440 \times 440 \times 9 \mu$m$^{3}$ in about 8 hours. The reconstruction revealed intra-grain misorientations exceeding \(10^{\circ}\) (see fig. S7) along with stress localization on several length scales. Such comprehensive data offer a unique avenue for guiding, optimizing, and validating multi-length scale models. The data can be interfaced directly to voxelated simulation models (similar to, for example \cite{Zhang2020}), or statistical correlations between microstructural and stress variables can be extracted from the extensive dataset.

\begin{figure}[!ht]
    \centering
    \includegraphics[scale=1.0]{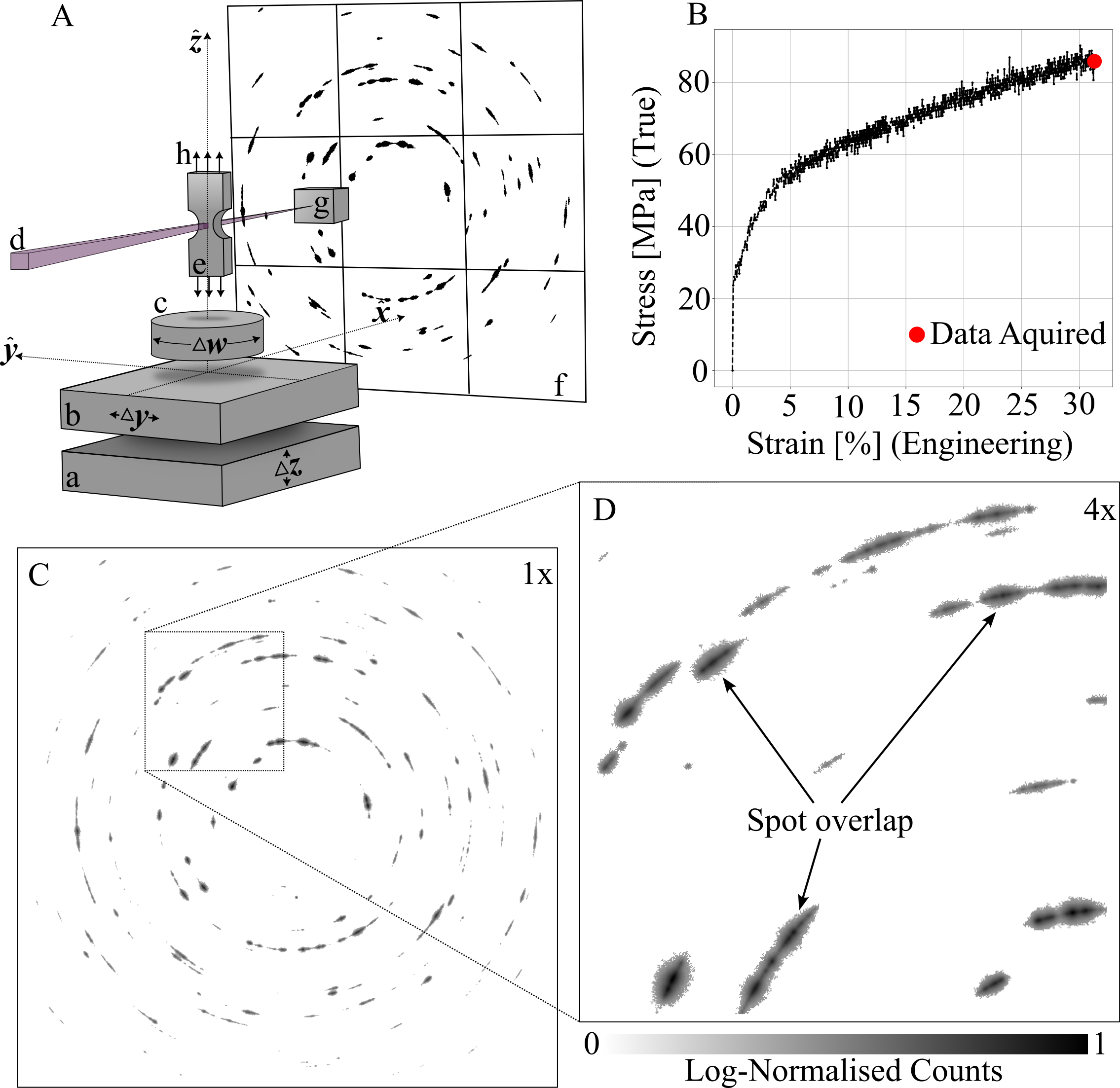}
    \caption{\textbf{Experimental principle} (A) Schematic of the X-ray diffraction imaging setup.  The sample is scanned in directions \(\boldsymbol{\hat{y}}\) and \(\boldsymbol{\hat{z}}\) (a,b) and rotated (c) while probed with a focused X-ray beam (d). The diffraction signal from the notched sample (e) is acquired by an area detector (f), (g) is a beamstop and (h) illustrates the  tensile load pattern. (B) The stress-strain curve  with indication of the ultimate tensile strength (\(\sim 85\) MPa) where imaging was performed. (C) Diffraction pattern acquired while integrating over a \(\Delta \omega=1^{\circ}\) sample rotation interval. (D) Zoom-in. Plastic deformations gave rise to peak arcing around the diffraction rings (C) and diffraction spot overlap (D).}
    \label{fig:method}
\end{figure}

The setup is illustrated schematically in Fig. \ref{fig:method}A. The sample was notched to enable mapping at the precise location of the macroscopic stress localization. Following \emph{in situ} tensile deformation, cf. Figs. \ref{fig:method}B and S2, the sample remained under load and underwent tomographic raster scanning in steps of 3\(\mu\)m by moving a \(~450 \times 450 \times  9 \mu \)m\(^3\) volume of interest across a 3\(\mu\)m \(\times\) 3\(\mu\)m X-ray microbeam while rotating the diffractometer. The plastic deformation at this deformation level is evident through substantial spot overlap in the diffraction images and peak spreading around the diffraction rings, as depicted in Fig. \ref{fig:method} C,D. For further details on the materials and X-ray methodology, see Supplementary Information.

 Using algorithms from computed tomography, the illuminated sample volume was partitioned into 247,956 voxels (see fig. S3) \cite{SupplementaryMaterials}. For each voxel, crystal orientations and strains consistent with the diffraction data were determined using ImageD11 \cite{ID11} and algorithms similar to those presented in \cite{Moscicki2009,Oddershede2010,Edmiston2011,Sharma2012,Schmidt2014,Hayashi2015,Kim2023,Kim2023Inclined}. This resulted in an orientation-strain map which listed the best candidate crystal lattices found for each voxel in the sample, with each voxel being treated independently of its neighbors. These local orientations and strains represent a multi-channel sampling of possible solutions to the inversion problem of retrieving the microstructure from the data. Due to the high dimension of the solution space, any approach reliant on trial-and-error optimization becomes computationally impossible.

For small amounts of deformation, where diffraction spot overlap remains limited, grain shapes can be reconstructed from diffracted intensities using tomographic methods \cite{Poulsen2003FBP,Poulsen2003ART,Xiaowei2006, Alpers2006,Rodek2007,Bonnin2014}. Subsequently, once the grain shapes are determined, the strain-orientation field within individual grains can be inferred by solving another, modified, tomographic reconstruction problem \cite{Henningsson2023}. However, for industrially relevant plastic deformation, diffraction peaks broaden and exhibit multiple local maxima, rendering tomographic grain shape reconstruction unfeasible.

In addition to tomographic reconstruction approaches, point-wise fitting methods have emerged, avoiding the use of diffracted intensities \cite{Hayashi2015,Hayashi2017,Hayashi2019,Kim2023, Hayashi2023ModCompletness}. These methods reconstruct a strain-orientation map by selecting, for each voxel, the orientation with the highest completeness (i.e. the ratio of the number of observed diffraction peaks to expected peaks). While these methods theoretically produce a strain-orientation map at various deformation levels, the presence of erroneously scattered orientations increases with diffraction peak overlap. Consequently, distinguishing among solutions within the multi-channel orientation-strain map becomes challenging, particularly at elevated levels of plastic deformation. 

To address these challenges, we introduce a new reconstruction algorithm. Our framework incorporates \emph{a priori} knowledge of the spatial correlation of orientations by employing median filters directly on inverse pole figure space. By maximizing completeness and simultaneously enforcing spatial correlation between the orientations of neighboring voxels, we overcome the challenges posed by diffraction spot overlap. Consequently, we can efficiently search the multi-channel orientation-strain map for a solution, even at industrially relevant degrees of plastic deformation. We implemented our algorithm separately for each \(z\)-layer of the volume and then stacked the solutions to construct a 3D grain volume (see \cite{SupplementaryMaterials} for algorithm details).

\begin{figure}[!ht]
    \centering
    \includegraphics[scale=1.0]{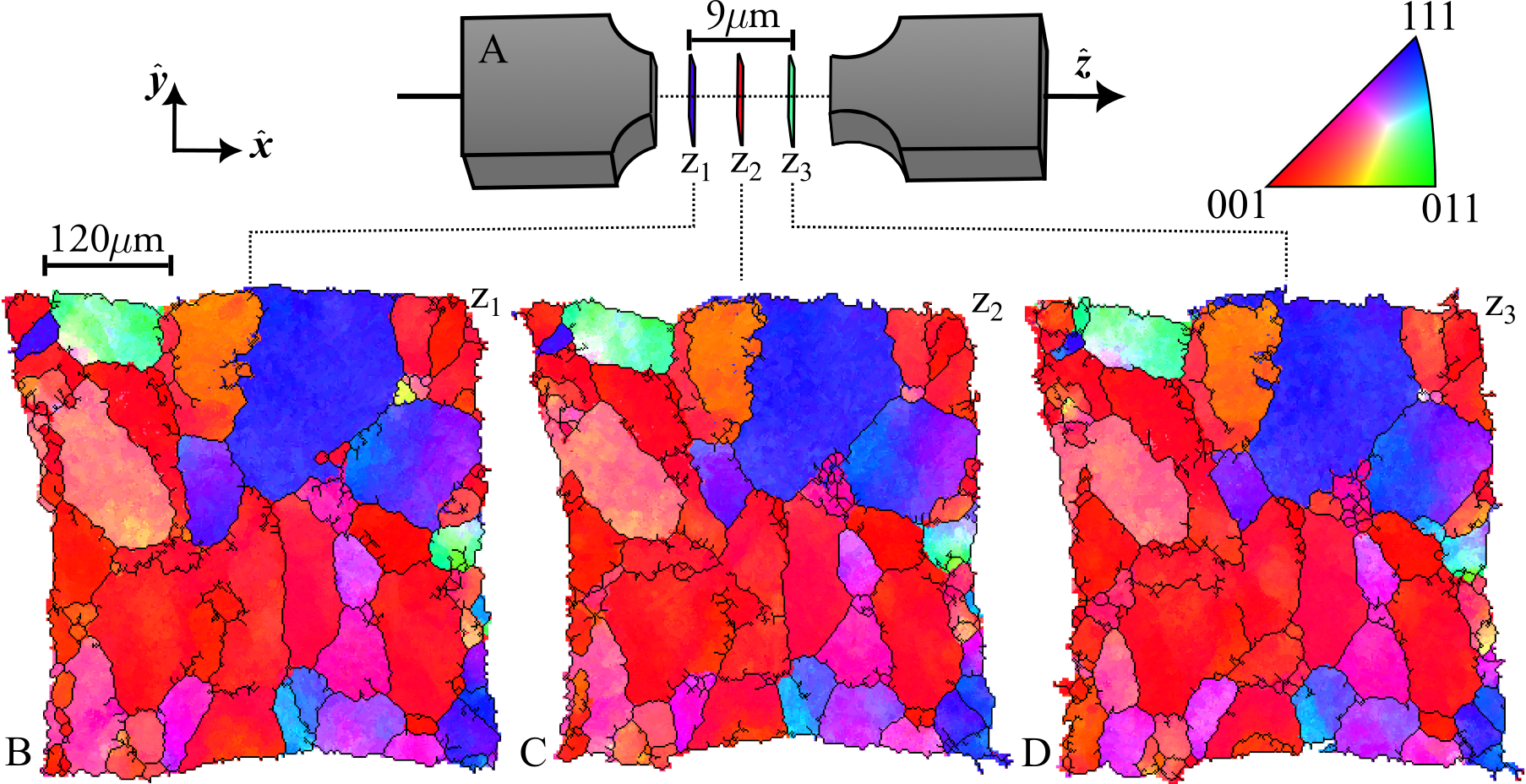}
    \caption{ \textbf{Reconstructed orientation maps.} (A) Schematic of the aluminum sample highlighting the three slices (z1, z2 and z3) scanned, evenly spaced by 3 \(\mu\)m in \(z\) and located at the sample notch center. (B-D) The orientation maps. The color map (top right) indicates the inverse pole figure (IPF) illustrating which crystallographic plane normal aligns with the tensile loading axis (\(\boldsymbol{\hat{z}}\)). The maps have been overlaid with grain boundaries (black lines) defined by a local misorientation threshold of $4 ^{o}$. }
    \label{fig:ipf_maps}
\end{figure}

The reconstructed orientation field is shown in Fig. \ref{fig:ipf_maps}. From this grain and sub-grain boundaries were identified by setting thresholds on the local mis-orientation (see figs. S4-S5) \cite{SupplementaryMaterials}. A plot of the average orientations of the resulting 60 grains (fig. S6) revealed the absence of grains with the tensile axis (\(\boldsymbol{\hat{z}}\)) oriented along the crystallographic (011) direction - as expected after tensile deformation. Intra-grain misorientations reaching up to 10\(^{\circ}\) were identified (fig. S7) which is to be compared to misorientations in the range of 1.5-3\(^{\circ}\) reported in previous S3DXRD studies \cite{Li2023}\cite{Hayashi2019}. 
In the supplementary materials \cite{SupplementaryMaterials}, a model is presented for the direct estimation of experimental error on the position of the grain boundaries. The standard deviation is found to be less than the beam size. 

\begin{figure}[!ht]
    \centering
    \includegraphics{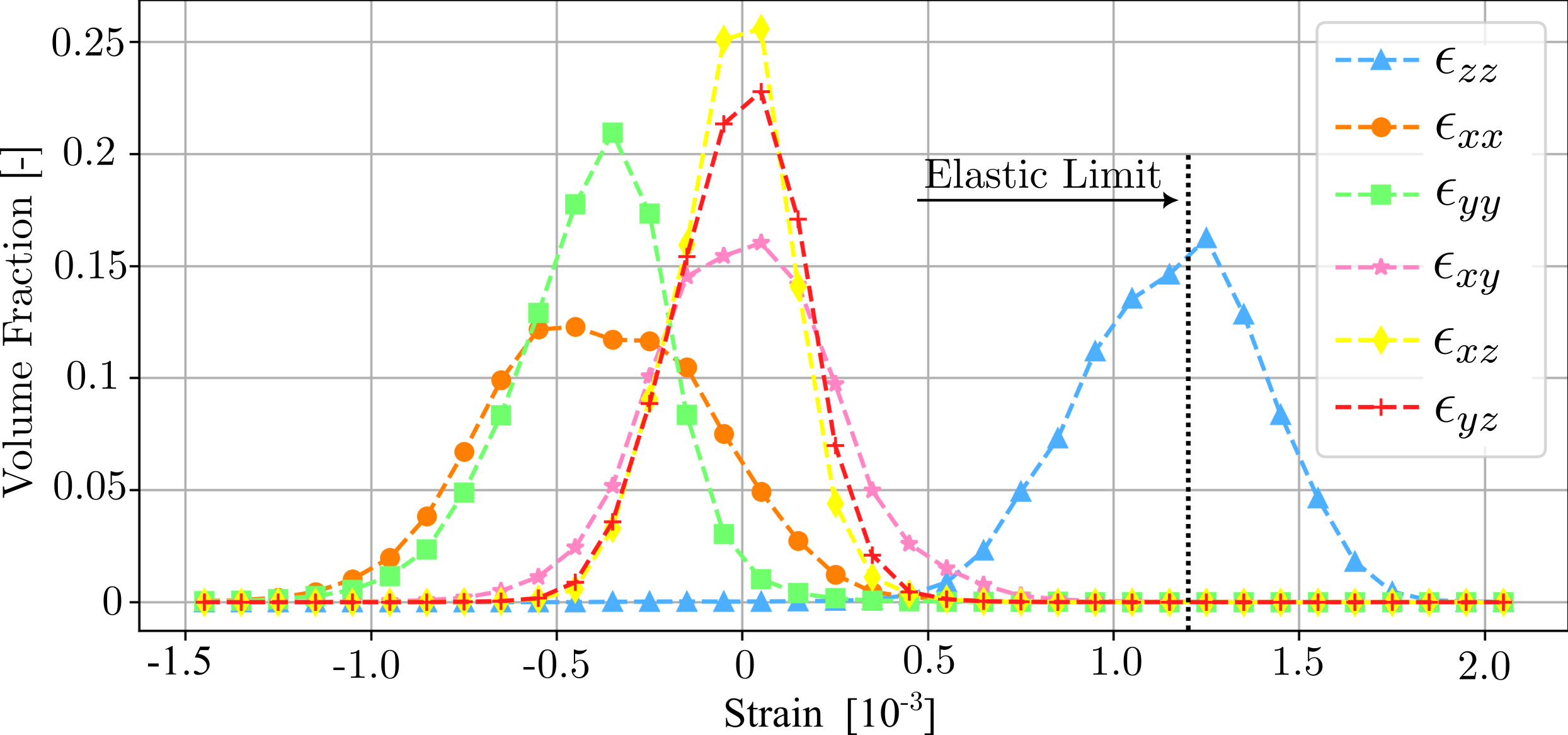}
    \caption{\textbf{Distributions of the six independent components of the elastic strain tensor.} Statistics over the 247,956 voxels in the sample volume. The axial strains along the tensile loading direction, \(\epsilon_{zz}\), are distributed with a mean strain close to the elastic limit of the alloy. }
    \label{fig:strain_dist}
\end{figure}

We simultaneously reconstructed the elastic strain tensor field across the same volume (see figs. S8-S10). From these maps, we derived statistics on the distributions of intra-grain strain components, as illustrated in Fig. \ref{fig:strain_dist}. 
The mean strain along the tensile loading axis (Fig. \ref{fig:strain_dist}, \(\epsilon_{zz}\)) is 0.0011, to be compared to an elastic limit of 0.0012 (considering a proof stress of 85 MPa and an average tensile Young's Modulus of 70 GPa). Remarkably, this implies that 42\% of the volume is strained above the limit. The ratio between the mean transversal strains and the mean axial strain yields a Poisson ratio in the range 0.34-0.36, to be compared to a literature value of 0.33. Furthermore, the shear strain components \(\epsilon_{xy}\), \(\epsilon_{xz}\) and \(\epsilon_{yz}\) have mean values of zero, as expected for uniaxial tension. 

\begin{figure}[!ht]
    \centering
    \includegraphics[scale=0.9]{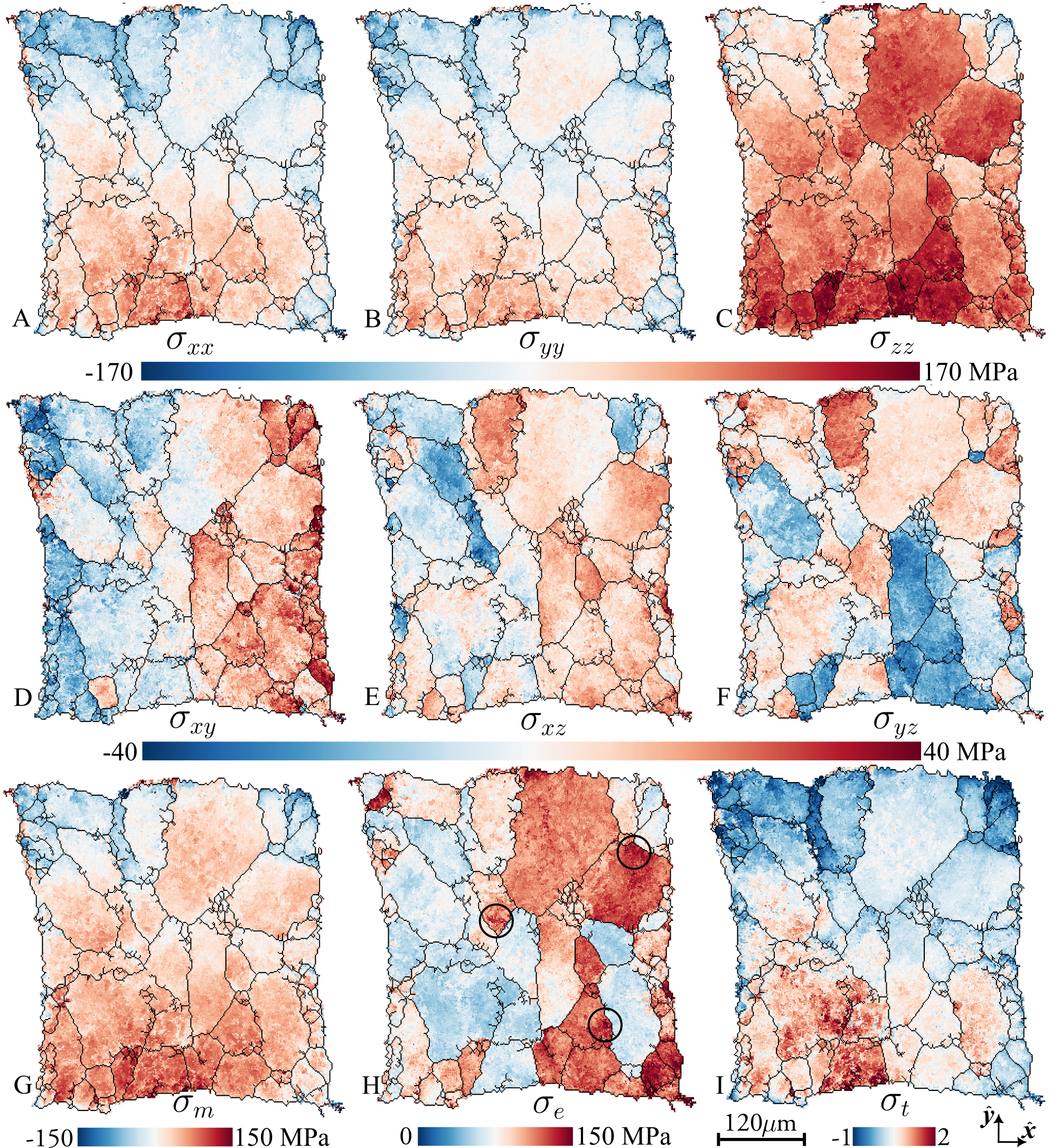}
    \caption{\textbf{Reconstructed stress tensor field.} (A-C) axial stress components \(\boldsymbol{\sigma}_{xx}\), \(\boldsymbol{\sigma}_{yy}\) and \(\boldsymbol{\sigma}_{zz}\), respectively.  (D-F)  Shear stress components \(\boldsymbol{\sigma}_{xy}\), \(\boldsymbol{\sigma}_{xz}\) and \(\boldsymbol{\sigma}_{yz}\), respectively. (G-I) Stress measures, derived from the stress tensor field: (G) the hydrostatic stress, \(\boldsymbol{\sigma}_{m}\), (H) the equivalent tensile stress, \(\boldsymbol{\sigma}_{e}\), and (I) the stress triaxiality, \(\sigma_t = \sigma_m / \sigma_e\). Three local intra-grain stress concentrations have been marked by solid circles in (H). All maps relate to the central layer z2. }
    \label{fig:stress_maps}
\end{figure}

Using  Hooke's law \cite{SupplementaryMaterials} and elastic constants for single crystal aluminum \cite{deJong2015}, the strain field was converted to a corresponding stress field. The resulting maps of the stress components are shown in Fig. \ref{fig:stress_maps}A-F (complemented by figs. S11-S13). Additionally, we present three derived measures. The local equivalent tensile stress, \(\sigma_{e}\), Fig. \ref{fig:stress_maps}H, is a measure of local distortion strain energy, commonly employed in predicting the onset of plastic deformation. Notably, this measure exceeds the true macroscopic flow stress at 32\% elongation (\(\sim 85 \)MPa) in \(\sim\)49\% of the volume. The hydrostatic stress, \(\sigma_{m}\), and the stress triaxiality, \(\sigma_t = \sigma_m / \sigma_e\), shown in Fig. \ref{fig:stress_maps}G and I, respectively,  are crucial metrics \cite{Andrade2016} in ductile failure theory.  Elevated values tend to promote ductile failure through void expansion and coalescence. An analysis of the stress resolution based on out-of-balance forces is given in the supplementary materials (see fig. S17). The standard deviations of the six (\(\sigma_{xx}, \sigma_{yy}, \sigma_{zz}, \sigma_{xy}, \sigma_{xz}, \sigma_{yz}\)) stress tensor components for each voxel were found to be in the range 8-12 MPa. This noise level is sufficient for testing most modeling approaches.

Inspection of Fig. \ref{fig:stress_maps} reveals a heterogeneous distribution of stress across various length scales, namely within the sample and at the inter-grain and the intra-grain levels. On the sample scale, the axial stress plots in Fig. \ref{fig:stress_maps}A-C exhibit high values in the lower left corner, consequently resulting in high hydrostatic stress and triaxiality in Fig. \ref{fig:stress_maps}G and I. This observation is attributed to the onset of necking in the sample, altering the shape of the initially square cross-section. 

\begin{figure}[!ht]
    \centering
    \includegraphics[scale=1.0]{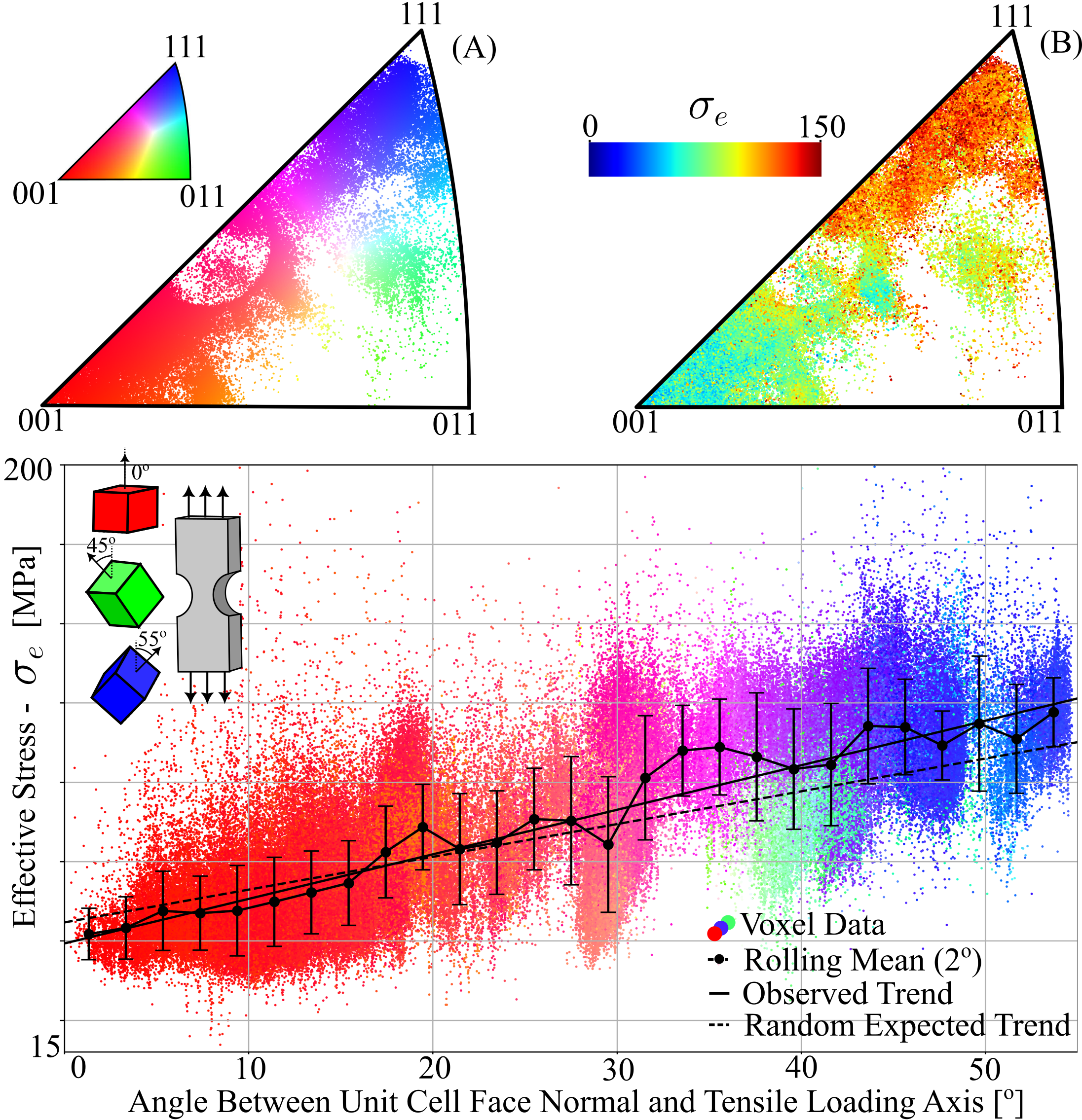}
    \caption{\textbf{Orientation dependence of the effective stress.} (A) IPF orientation distribution of the 247,956 voxels in the sample volume as viewed from the tensile loading axis. Fully saturated red corresponds to a local cubic unit cell with one of its six facets at a 90$^{\circ}$ angle to the loading axis, as illustrated in the top left corner of subfigure (C). (B) Corresponding local equivalent tensile stress, \(\sigma_e\). (C) Relationship between \(\sigma_e\) and the angle between the cubic unit cell faces and tensile loading axis. The scatter points are colored by the IPF color key in (A) and the error bars on the rolling mean curve mark one standard deviation. A linear least squares fit results in a slope of \( \sim 1.4 \) MPa / degree. In comparison, a simple model considering only the anisotropic elastic moduli of aluminum gives a slope of \( \sim 1.0 \) MPa / degree.}
    \label{fig:effective_stress_trend}
\end{figure}

A distinction is also observed between the left and right parts of Fig. \ref{fig:stress_maps}D-F and H. In particular, it is noteworthy that $\sigma_e$ primarily exceeds the macroscopic flow stress in the right part of the sample. A qualitative comparison with the crystallographic direction of the tensile axis in Fig. \ref{fig:ipf_maps}C reveals that grains featuring unit cell face normals aligned with the loading direction exhibit lower than average $\sigma_e$ values. Similar orientation effects of grain averaged stress right after the onset of plastic deformation have also been observed \cite{Juul2017}. 
The influence of the grain orientation was quantified by calculating the angle between the loading axis and the unit cell face normal for all voxels. A near linear relationship between this angle and the equivalent tensile stress (Fig. \ref{fig:effective_stress_trend}C, solid line) was observed. This can partly be explained by the anisotropic stiffness of aluminum (dashed line).  The residual difference in slope indicates systematic stress partitioning between (111) domains and (001) domains due to crystallography. This is attributed to the increased hardening of these orientations due to the higher plastic work required for elongating the grain.  The clustering of these domains in the left and right part of the sample is the origin of the macroscopic partitioning, demonstrating the relevance of incorporating clusters of grains in modeling.

The linear trend of Fig. \ref{fig:effective_stress_trend}C predicts a stress fluctuation of 76 MPa over the sample volume. This stress variation is dominated by differences in inter-grain orientation and serves to measure the inter-grain stress heterogeneity in the sample (referred to as type II stress). In contrast, the distribution of equivalent tensile stress in Figure \ref{fig:stress_maps} H reveals stress variations below the scale of individual grains. These intra-grain stress variations, classified as type III stress, influence the extent of the error bars in Fig. \ref{fig:effective_stress_trend}C. To quantify the effective type III stress variation, we converted the error bars in Fig. \ref{fig:effective_stress_trend}C into equivalent full width half maxima. This conversion allowed us to establish an effective type III stress, yielding a value of 36 MPa. Consequently, our analysis indicates that the inter-grain stress range spans approximately twice the magnitude of the intra-grain stress range (76 MPa / 36 MPa). This finding underscores the importance of multi-scale modeling.

These three examples illustrate how local stress maps can be incorporated into damage and flow models, respectively. As another example, we propose that the scatter plot data presented in Fig. \ref{fig:effective_stress_trend}C can provide unique information for addressing the longstanding question regarding the significance of grain orientation versus grain interactions, and for quantifying intra-grain stress distributions as a function of grain orientation and the degree of deformation.

In our approach diffraction contrast imaging constitutes a spatial classification problem. In contrast to previous efforts which have generated orientation-strain maps based on a local maximum completeness criterion, our novel algorithm capitalizes on the concept of a multi-channel orientation-strain map encoding the set of possible solutions. By exploiting the spatial correlation in the orientation field we have devised a streamlined approach to reduce the vast solution set, that prioritizes fidelity to data while yielding a single stress-orientation field. The quality of the data is validated by our error analysis, figs. S14, S16-S17, and the correspondence with macroscopic properties, as shown e.g. in Fig. \ref{fig:strain_dist}. 

The experimental setup allows for simultaneous absorption contrast tomography, thereby establishing a direct link between microstructure and stress evolution and the nucleation and growth of voids and cracks. Our classification approach is scale-invariant and can be used in connection with S3DXRD studies of polycrystals featuring nano-meter sized grains, utilizing nano-beams as small as 100-250 nm \cite{Bonnin2014, shukla2024grain, Hektor2019}. Furthermore, there exist several avenues for significantly enhancing strain and stress accuracy. Firstly,  the advent of larger area detectors with more pixels will enable improved differentiation between spatial and strain degrees of freedom \cite{Kutsal2022}. Secondly, our approach classifies the multi-channel orientation-strain map without using scattered intensity in the forward diffraction model. Including the intensities in the future is expected to yield substantial improvements in reconstruction quality. As an example, in 3DXRD, Monte-Carlo based algorithms using Gibbs priors have shown promise \cite{Alpers2006}.

From a broader perspective, we envision that X-ray diffraction microscopy can address a range of key questions in metallurgy by characterizing industrially relevant degrees of plastic deformation in 3D and in-situ. The capability to track deformation processes in three dimensions, spanning from the initial elastic regime through the plastic regime to component failure, will be pivotal moving forward. This capability holds the potential to bridge the gap between nano- and macroscale mechanics, addressing inquiries in predictive materials modeling on fracture phenomena such as creep, fatigue, and ductile failure theory. 
Furthermore, modern metalworking industries stand to benefit significantly from this advancement, enhancing their ongoing efforts to customize extrusion, machining, fabrication, and rolling processes. The optimization of these processes for material property enhancement offers sustainability and recycling advantages compared to alloying with additional chemical elements.

\bibliography{scibib}
\bibliographystyle{Science}

\section*{Acknowledgments}

We acknowledge ESRF for the beamtime (MA4756). The computations were enabled by resources provided by LUNARC, The Centre for Scientific and Technical Computing at Lund University.

\textbf{Funding:} 
This work was funded by the Independent Research Fund Denmark (grant no 0136-00194) (GW), the European Research Council Advanced grant 885022 (HFP), Vetenskapsrådet - Röntgen Ångström Cluster project number 2017-06719 (SH), the Danish Agency for Science and Higher Education grants number 8144-00002B (HFP) and 7129-00006B (HFP) and DanScatt for a travel grant. 

\textbf{Author contributions:}
The manuscript was primarily authored by AH, with contributions from HFP and GW, and subsequent edits from all co-authors. The materials science concept of the study was developed by GW and HFP, while AH conceived and implemented the new algorithm under the supervision of SAH and JPW. MK, JPW, and WL led the planning and execution of the experiment, with contributions from HFP, GW, and HOS. AH conducted the data analysis, with supplementary analyses by MK, JPW and WL.

\textbf{Competing Interests:} All authors declare no competing interests.

\textbf{Data and materials availability:}  All data, code, and materials used in the analysis will be provided to any researcher to reproduce or extend this analysis. The original synchrotron data and metadata is available at \newline
\url{https://data.esrf.fr/doi/10.15151/ESRF-ES-578179478}.
The python codes used to analyze the data, the segmented diffraction peak files, and the processed results are available at \url{https://zenodo.org/doi/10.5281/zenodo.11058847}.

\newpage

\baselineskip12pt
\renewcommand{\thefigure}{S\arabic{figure}}
\renewcommand{\thetable}{S\arabic{table}}
\setcounter{figure}{0}

\section*{Supplementary Materials}

\section{Materials and Methods}

\subsection{Sample description}
\label{sec-sample}

The sample material was AA1050, which was cold-rolled and subsequently recrystallized, resulting in an average grain size of 50-70\(\mu\)m and a typical recrystallization texture characterized by the predominance of the Cube orientation. This orientation signifies the alignment of the axes of the crystallographic unit cell with the rolling coordinate system. 

The sample was cut such that the tensile axis was inclined at 45\(^\circ\) to the rolling direction. Consequently, the specimen initially exhibited a prevalence of grains with the (011) plane normal aligned with the tensile axis and the (001) plane normal aligned with the X-ray beam direction.

\begin{figure}[H]
    \centering
    \includegraphics[scale=2.0]{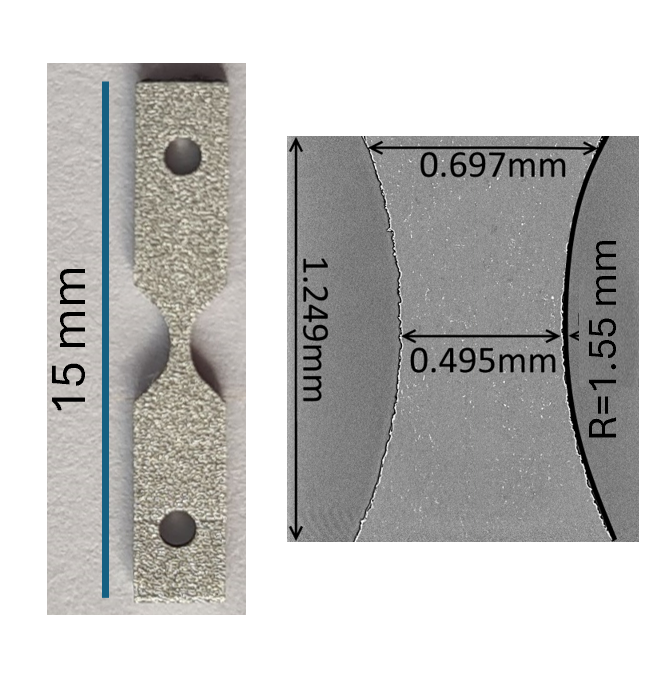}
    \caption{\textbf{Tensile specimen}. Left: Photo of the sample. Right: Tomographic slice of the central region. S3DXRD was conducted where the sample had the smallest cross-sectional area. }
    \label{fig:sampleshape2}
\end{figure}

The central part of the sample was notched with a radius of 1.55 mm and a square cross-section at the thinnest region measuring 0.24 mm\(^2\) (see fig. \ref{fig:sampleshape2}). The total length of the notched region was 1.25 mm.

\subsection{Experimental setup}
Diffraction from the tensile specimen was imaged by scanning 3DXRD at the ID11 beamline of ESRF using the 3DXRD station at an X-ray energy of 42.5 keV. The sample was raster scanned in steps of 3\(\mu\)m by moving a \(~450 \times 450 \times  9 \mu \) m\(^3\) volume of interest across a 3 \(\mu\)m \(\times\) 3 \(\mu\)m X-ray microbeam. At each scan position, the sample was rotated over 180\(^\circ\), and digital 2D (2K, 16-bit) images were acquired by integrating the diffraction signal over 1.0\(^\circ\) intervals using a charge coupled device (CCD) FReLoN detector \cite{Labiche2007} placed 98.9 mm downstream of the sample with pixel dimensions of 47.2 \(\times\) 47.2 \(\mu\)m\(^2\). Following the density within the central part of the notched region of the sample was mapped by attenuation contrast imaging.

\subsection{Stress rig}
\begin{figure}[H]
    \centering
    \includegraphics[scale=1.0]{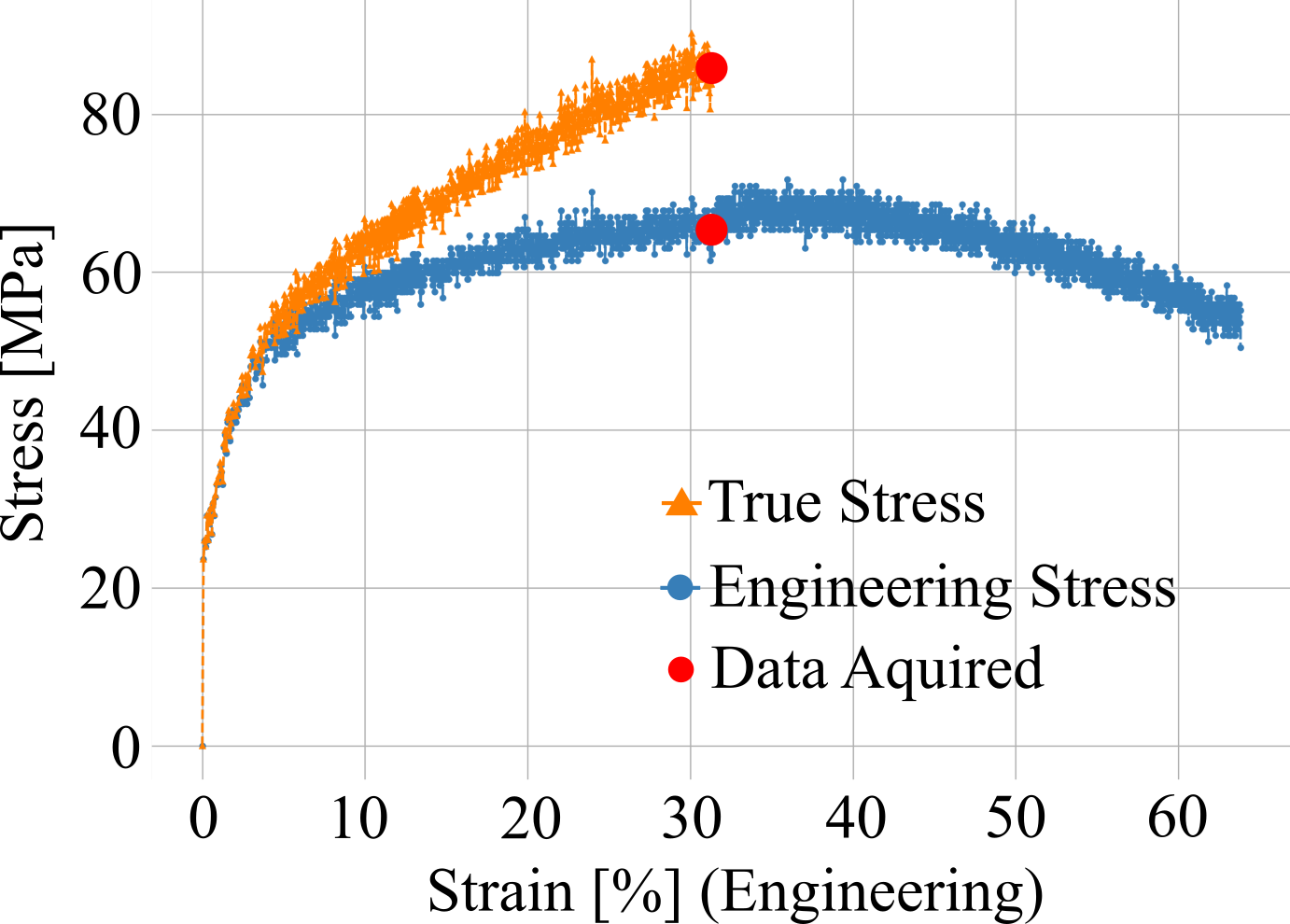}
    \caption{\textbf{Macroscopic stress-strain response}. The tensile stresses in the notched aluminium sample are plotted against tensile engineering strains. The point at which S3DXRD diffraction data was collected is indicated by a red dot.}
    \label{fig:stress_vs_strain}
\end{figure}

The sample was mounted in the Nanox stress rig, which was provided by the ESRF \cite{Gueninchault2016}. Tensile deformation was then applied \emph{in situ}. Both the engineering stress and the true stress (corrected for variable cross-sectional area) are shown in fig. \ref{fig:stress_vs_strain}. The S3DXRD study was conducted at approximately \(65\) MPa of engineering stress, which corresponds to the ultimate tensile strength of the alloy.
Due to the sample shape, elongation is heterogeneous, and the local elongation at the location of the S3DXRD scans was determined to be 32\% based on the reduction of the cross-sectional area (with consistent results provided by s3DXRD mapping and tomography). This result was also confirmed by tracking, in a similar sample, two pre-existing voids. The voids were initially situated 150 \(\mu\)m apart and located just below and above the S3DXRD region of interest. X-ray attenuation contrast tomography was used to track their position with deformation (results to be published elsewhere).

\section{Methodology and data analysis}

\subsection{Orientation-strain map reconstruction algorithm}
\label{sec-orientation-strain-reconstruction}
The data analysis pipeline used to reconstruct the voxelated orientation-strain map is delineated by the following six algorithmic steps (I-VI). Our algorithm was applied on a layer-by-layer basis, where each of the three layers (z1, z2, z3) was reconstructed independently. These reconstructed layers were then stacked to obtain a 3D reconstructed volume. The Python implementation of the following algorithm is openly available at \url{https://zenodo.org/doi/10.5281/zenodo.11058847}. While the computationally intensive parts of our algorithm were optimized for high-performance computing clusters, the code can be deployed on most CPU-based systems.

\subsubsection*{Step I - Peak segmentation}
The diffraction patterns (\(\sim10^5\) 2K 16-bit images) underwent  background correction, and the signal was segmented to produce sparse images where only intense signals (\>25 ADU) composed of connected pixels (with at least 9 pixels forming a spot) were retained. To label the connected pixels as diffraction spots we start by assigning each pixel to a gradient vector. The gradient vector of a pixel is here defined as the 2D vector that connects the pixel with its locally maximal neighbor pixel. Secondly, we assign unique labels to all pixels that constitute a local intensity maximum. Finally, we back-propagate the unique labels through the paths defined by the gradient vectors. Consequently, patches of connected pixels featuring multiple local maxima will be segmented into multiple distinct diffraction peaks. Using this method a list of \(\sim\)14,000,000 diffraction peaks were recorded, each associated with a detector position centroid, \(\boldsymbol{r}\), a turntable rotation angle, \(\omega\), a sample translation, \(\Delta y\), and an integrated intensity count.

\subsubsection*{Step II - Sample mask reconstruction}
\begin{figure}[H]
    \centering
    \includegraphics[scale=0.9]{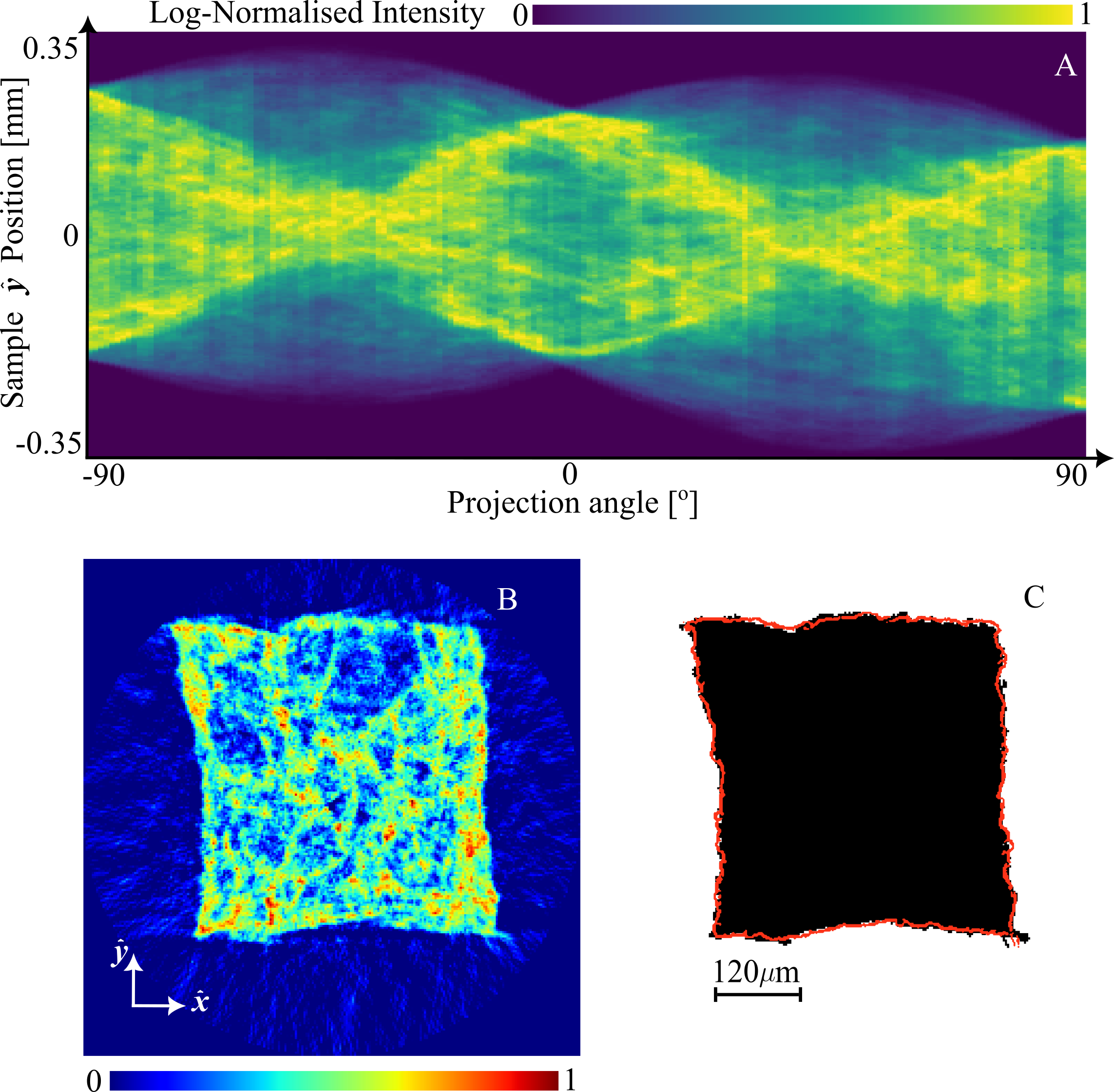}
    \caption{\textbf{Sample cross-section reconstruction}. The log-normalised sinogram representing the summed diffracted intensity (A) was subjected to filtered back-projection (B), facilitating the segmentation of the sample cross-section shape (C). The outline of the sample obtained independently from absorption tomography is highlighted in red. The figure corresponds to the central layer (z2).}
    \label{fig:sampleshape}
\end{figure}

We utilized the summed diffraction signal on each collected detector frame to construct a sinogram, cf. fig. \ref{fig:sampleshape} A. The collected intensity was log-normalized per column in the sinogram, and the inverse radon transform (filtered back-projection) was applied. This process was implemented using a linear interpolation scheme and a standard ramp filter, as provided by the scipy library in Python \cite{scipy}. 

The back-projected intensity for the central layer (z2) can be seen in fig. \ref{fig:sampleshape} B. The sample shape mask, depicted in black in fig. \ref{fig:sampleshape} C, was extracted from the back-projected reconstruction through the following steps: (I) thresholding the normalized reconstruction at 0.2, (II) selecting the largest connected domain in the thresholded reconstruction, and (III) filling all holes on the selected domain. The overlaid sample outline in red (fig. \ref{fig:sampleshape} C) was obtained by absorption tomography of the same layer in the sample.

Using a voxel size of \(1.5\) \(\times\) \(1.5\) \(\times\) \(3\mu\) m\(^3\), the number of voxels occupied by material was found to be 247,956. The advantages of selecting the \(x\)-\(y\) voxel dimensions smaller than the beam size are discussed elsewhere \cite{Kim2023, Kim2023Inclined}.

\subsubsection*{Step III - Indexing}
We indexed candidate crystal orientations at each voxel in the volume using ImageD11 \cite{ID11} which implements indexing algorithms similar to those described in \cite{Moscicki2009, Oddershede2010, Edmiston2011, Sharma2012, Schmidt2014}. By compiling a local diffraction data set composed of diffraction events where the distance between the X-ray beam and voxel centroid less than the beam size (3\(\mu\)m), candidate orientation-strain states were independently indexed for each voxel. 

Explicitly, the voxel centroid position, \(x,y\), given in a fixed laboratory coordinate system (see \cite{Henningsson2020} for coordinate system formalism) is a function of the turntable rotation angle, \(\omega\), as
\begin{equation}
\begin{split}
    x = x_0  \cos(\omega) - y_0  \sin(\omega) \\
    y = x_0  \sin(\omega) + y_0  \cos(\omega) + \Delta y,
\end{split}
\label{eq:rotation}
\end{equation}
where (\(x_0,y_0\)) is the position of the voxel at \(\omega=0\) and \(\Delta y\) is the sample translation in \(y\). The diffraction data associated with a specific voxel is defined to satisfy\(|y| \leq 3\mu\text{m}\). 

The sample diameter (0.24 mm) was sufficiently large to induce measurable shifts in the high-angle diffraction peak positions (with a sample-to-detector distance of 98.9 mm). Consequently, the diffraction data associated with the voxels were calibrated to account for the voxel position along the beam path (\(x\)). This entailed computing (elastic) scattered wave vectors, \(\boldsymbol{k'}\), as
\begin{equation}
    \boldsymbol{k'} = \dfrac{2\pi}{\lambda}\dfrac{\boldsymbol{r} - \boldsymbol{x}}{\sqrt{(\boldsymbol{r} - \boldsymbol{x})^T(\boldsymbol{r} - \boldsymbol{x})}},
    \label{eq:diffraction_origins}
\end{equation}
where \(\boldsymbol{r}\) is the coordinate of a diffraction peak in 3D laboratory coordinates, and \(\boldsymbol{x}\) is the voxel position according to equation \eqref{eq:rotation}. Diffraction vectors were defined as
\begin{equation}
    \boldsymbol{G}_l = \boldsymbol{k'} - \boldsymbol{k},
\end{equation}
where \(\boldsymbol{k}\) is the incident wave vector (which is parallel to \(\boldsymbol{\hat{x}}\) in our setup). The corrected set of diffraction vectors, \(\boldsymbol{G}^{(1)}_l, \boldsymbol{G}^{(2)}_l, ..., \boldsymbol{G}^{(m)}_l\), belonging to the voxel, was input to the indexer of ImageD11. The number of successfully indexed strain-orientation states found for each voxel depends on a minimum threshold of the number of peaks that are indexed within a given \(h,k,l\) error tolerance (see step IV and equation \eqref{eq:hkl_tol}). 

With a very low \(h,k,l\) error tolerance the ImageD11 indexing algorithm may output many similar orientations, while a very high \(h,k,l\) error tolerance results in some peaks being assigned incorrectly. The values chosen here (0.03-0.05) correspond to an angular misorientation of 1-1.65 \(^\circ\) for the (111) ring and 0.3-0.5\(^\circ\) for the (600) ring which corresponds to the highest angle measured. The (311) ring was found to be sufficient for initial indexing.

We relaxed the \(h,k,l\) error tolerance linearly from 0.03 to 0.05 until at least two distinct orientations could be indexed for the considered voxel. Similarly, the minimum number of diffraction peaks required to be indexed by an orientation was linearly relaxed from 216 to 108 alongside the \(h,k,l\) error tolerance. Additionally, the maximum number of candidate orientation-strain states allowed to be indexed by a single voxel was capped at 10.

\subsubsection*{Step IV - Orientation and strain refinement }
We merged all diffraction peaks across detector frames, ensuring that each diffraction event could be associated to a unique set of Miller indices, \(h,k,l\), along with a sample translation, \(\Delta y\). Subsequently, for each voxel, we refined all candidate orientation matrices and strain tensors using all 13 available diffraction rings.

To describe the refinement procedure, we introduce the unit cell matrix, \((\boldsymbol{UB})^{-T}\), which contains the lattice unit cell vectors as its columns. The unit cell matrix is related to the diffraction vector, \(\boldsymbol{G}_l\), according to the Laue equations
\begin{equation}
    \boldsymbol{G}_l = \boldsymbol{\Omega}\boldsymbol{UB}\boldsymbol{G}_{hkl},
\end{equation}
where \(\boldsymbol{\Omega}\) is a rotation matrix describing the sample rotation with the turntable and \(\boldsymbol{G}_{hkl}\) are integer Miller indices (c.f \cite{Poulsen2004} for details on the diffraction formalism). For simplicity we refer to \(\boldsymbol{G}\) in sample coordinates, such that
\begin{equation}
    \boldsymbol{G} = \boldsymbol{UB}\boldsymbol{G}_{hkl}.
    \label{eq:laue}
\end{equation}
In step III, we utilize equation \eqref{eq:rotation} to allocate a sub-set of diffraction vectors, \(\boldsymbol{G}^{(1)}, \boldsymbol{G}^{(2)}, ..., \boldsymbol{G}^{(m)}\), to the voxel under refinement. Since each diffraction vector is linked to both a turntable translation, \(\Delta y\), and rotation, \(\omega\), the area of intersection between the X-ray beam and the voxel becomes variable. Furthermore, the beam profile deviates from a perfect top-hat function, exhibiting tails that extend beyond the nominal cross section of 3\(\mu\)m \(\times\) 3\(\mu\)m. To accommodate these effects we introduce a data weight, \(w\), which remains non-zero over a range extending beyond the nominal beam size. Importantly, the data weight should diminish with increasing distance between the X-ray beam and the centroid of the voxel. We selected the data weights as
\begin{equation}
    w  = \dfrac{1\mu\text{m}}{( y + \Delta y + 1\mu\text{m})}.
    \label{eq:beam_profile_w}
\end{equation}
Consequently, when the voxel is centered on the beam (\(y+\Delta y=0\)), we find \(w=1\), and when the voxel has a maximal offset from the beam (\(y+\Delta y=\)3\(\mu\)m), we find \(w=0.25\).

To refine the unit cell matrix, \((\boldsymbol{UB})^{-T}\), we introduce the data vector 
\begin{equation}
    \boldsymbol{\Bar{\boldsymbol{G}}} = \begin{bmatrix}
        \boldsymbol{G}^{(1)}\\
        \boldsymbol{G}^{(2)}\\
        \vdots\\
        \boldsymbol{G}^{(m)}\\
    \end{bmatrix}
\end{equation}
where \(\boldsymbol{G}^{(i)}\) are diffraction vectors indexed by the candidate unit cell matrix, \((\boldsymbol{UB})^{-T}\), adhering to the \(h,k,l\) error tolerance, \(e_{hkl}=0.05\). Explicitly, the data satisfying
\begin{equation}
    ||(\boldsymbol{UB})^{-1}\boldsymbol{G}^{(i)}||_2 < e_{hkl}
    \label{eq:hkl_tol}
\end{equation}
was considered, where \(||\cdot||_2\) denotes the Euclidean norm. Furthermore, we introduce a flattened format of the unit cell matrix as
\begin{equation}
    \boldsymbol{\rho} = \begin{bmatrix}
        UB_{11}\\
        UB_{12}\\
        UB_{13}\\
        UB_{21}\\
        UB_{23}\\
        UB_{23}\\
        UB_{31}\\
        UB_{32}\\
        UB_{33} \\
    \end{bmatrix}
\end{equation}
allowing us to define a system matrix, \(\boldsymbol{H}\), populated by the integer Miller indices, \(h,k,l\), such that
\begin{equation}
    \boldsymbol{\Bar{\boldsymbol{G}}} = \boldsymbol{H}\boldsymbol{\rho}.
\end{equation}
We can now perform the least squares fit
\begin{equation}
    \boldsymbol{\rho} = (\boldsymbol{H}^T\boldsymbol{W}^T\boldsymbol{W}\boldsymbol{H})^{-1}\boldsymbol{H}^T\boldsymbol{W}^T\boldsymbol{W}\boldsymbol{\Bar{G}}
    \label{eq:lsq_ub}
\end{equation}
where \(\boldsymbol{W}\) is a diagonal weight matrix according to equation \eqref{eq:beam_profile_w}.

 The angular shift in a diffraction peak centroid can be attributed to a directional strain acting across the diffracting lattice planes \cite{Poulsen2001Strain}\cite{Allen1985}. Therefore, the precision in the elastic strain tensor (\(\boldsymbol{\epsilon}\)) fit can be further enhanced by directly exploiting these angular shifts in the diffraction data. Specifically, following \cite{Henningsson2021} each diffraction peak centroid was converted into a single scalar measurement of directional strain, \( \varepsilon\), as 
\begin{equation}
    \varepsilon = \dfrac{\boldsymbol{G}^T\boldsymbol{G}_{0}}{\boldsymbol{G}^T\boldsymbol{G}} - 1,
\end{equation}
where \(\boldsymbol{G}\) is the measured diffraction vector and \(\boldsymbol{G}_0\) is a model diffraction vector. The model diffraction vector is computed using the local fitted orientation matrix, \(\boldsymbol{U}\), and an undeformed reference unit cell, which is defined based on tabulated values (\(a=4.04\)Å, \(\alpha = 90.0^\circ\))\cite{Jette1935}.

Considering the symmetric elastic strain tensor, \(\boldsymbol{\epsilon}\), as the unknown parameter we defined the model equation
\begin{equation}
    \varepsilon = \dfrac{\boldsymbol{G}^T\boldsymbol{\epsilon} \boldsymbol{G}}{\boldsymbol{G}^T\boldsymbol{G}},
\end{equation}
which represents a strain tensor measurement in the direction normal to the lattice planes associated with \(\boldsymbol{G}\). Many such measurement form a linear set of equations
\begin{equation}
    \boldsymbol{y} = \boldsymbol{M} \boldsymbol{s},
\end{equation}
where \(\boldsymbol{s}\) holds the six unknown strain tensor components
\begin{equation}
\boldsymbol{y} = \begin{bmatrix}
        \varepsilon_{xx} \\
        \varepsilon_{yy} \\
        \varepsilon_{zz} \\
        \varepsilon_{xy} \\
        \varepsilon_{xz} \\
        \varepsilon_{yz}
    \end{bmatrix},
\end{equation}
and the rows of the system matrix, \(\boldsymbol{M}\), are defined as 
\begin{equation}
    M_j = \begin{bmatrix}
        \kappa^2_{1} &\kappa^2_{2} &\kappa^2_{3} &2\kappa_{1}\kappa_{2} &2\kappa_{1}\kappa_{3} & 2\kappa_{2} \kappa_{3},
    \end{bmatrix}
\end{equation}
and
\begin{equation}
    \boldsymbol{\kappa} = \begin{bmatrix}
        \kappa_1 \\ \kappa_2 \\ \kappa_3
    \end{bmatrix} = \dfrac{\boldsymbol{G}}{||\boldsymbol{G}||_2}.
\end{equation}
Assuming zero-mean isotropic Gaussian noise in \(\boldsymbol{G}\), with standard deviation \(\sigma_g = 10^{-4}\), we introduce the approximate weights
\begin{equation}
    w  = \bigg(\dfrac{1\mu\text{m}}{(y + \Delta y + \mu\text{m})}\bigg)\bigg(\sigma^2_g\dfrac{\boldsymbol{G}_{0}^T\boldsymbol{G}_{0}}{\boldsymbol{G}^T\boldsymbol{G}}\bigg)^{-1/2},
    \label{eq:strain_w}
\end{equation}
and define outliers as
\begin{equation}
    \varepsilon > \mu_{\varepsilon} + \text{3.5}\sigma_{\varepsilon} \quad \text{or} \quad \varepsilon < \mu_{\varepsilon} - \text{3.5}\sigma_{\varepsilon},
    \label{eq:outliers}
\end{equation}
where \(\mu_{\varepsilon}\) and \(\sigma_{\varepsilon}\) are the mean and standard deviation of the measured directional strains, respectively.

The refined strain tensor was determined using the weighted least squares solution
\begin{equation}
    \boldsymbol{s} = (\boldsymbol{M}^T\boldsymbol{W}^T\boldsymbol{W}\boldsymbol{M})^{-1}\boldsymbol{M}^T\boldsymbol{W}^T\boldsymbol{W}\boldsymbol{y}.
\end{equation}
where \(\boldsymbol{W}\) is a diagonal weight matrix according to equation \eqref{eq:strain_w}, and outliers have been disregarded according to equation \eqref{eq:outliers}.

This summarises our refinement step for the strain and orientation. The outlined procedure was executed independently for each voxel and each candidate orientation in the indexed volume.

\subsubsection*{Step V - Spatial filtering}
Combining the refined strain-orientation candidates with the sample mask resulted in a \(\sim 290 \times 290 \times 3\) voxel volume with a list of candidate grain orientations and strain tensors attached to each voxel. This multi-channel orientation-strain map represents a vast set of possible solutions to an inversion problem. On average, each voxel held 6 orientation channels, resulting in a huge number of possible solutions by randomly permuting choices. This complexity makes any approaches relying on trial-and-error selection computationally impossible.

The next step of our reconstruction approach aims to establish a single-channel orientation-strain map by selecting one solution from the multi-channel orientation-strain map. It is possible to use the maximum number of unique diffraction peaks (maximum completeness) as a selection criterion \cite{Hayashi2015, Hayashi2017, Hayashi2019, Hayashi2023ModCompletness}. However, due to the challenges of diffraction spot overlap, we found this criterion alone to result in noisy orientation-strain maps with artifact orientations scattered over the reconstructed volume. To overcome this obstacle, we aimed to preserve spatial correlation in the final orientation-strain map while maximizing the number of unique diffraction peaks in the reconstruction. This step in our reconstruction approach is summarized by the following 5 actions:
\begin{enumerate}
    \item We selected an initial (noisy) single-channel orientation-strain map by assigning, for each voxel, the candidate orientation that indexed the most (unique) diffraction peaks.
    \item We converted the noisy single-channel orientation-strain map into a color image (one RGB value per voxel) using an inverse pole figure color map.
    \item We ran a 6\(\times\)6 median filter across each color channel (Red, Green, Blue) in the image, resulting in a new, artificial color image with substantially reduced noise.
    \item We converted each channel in the original multi-channel map into RGB values again using the inverse pole figure color map.
    \item We selected a new single-channel orientation-strain map by comparing our artificial color image to the RBG values of the multi-channel map. Each voxel in the new single-channel map was assigned the candidate orientation-strain state from the multi-channel map that featured the smallest Euclidean norm between RGB tuples.
\end{enumerate}

\subsubsection*{Step VI - Diffraction centroid calibration}
In Step III of our analysis, we approximated diffraction to originate from the centroids of the individual voxels in the sample. This approximation can be improved by recognising that each diffraction peak is associated to a spatially extended domain in the sample. These diffracting domains can be approximated as intersections between the X-ray beam and the voxel grid, considering only voxels that fulfil the Laue equations \eqref{eq:laue}. Using the orientation-strain map obtained from step V, it is therefore possible to compute an updated approximation of the diffraction origins (\(\boldsymbol{x}\)). For each observed diffraction peak, we computed the centroid of the corresponding diffracting domain in the sample and inserted the result into equation \eqref{eq:diffraction_origins}, effectively correcting the diffraction vector data set. The analysis in step IV was then rerun, updating the multi-channel orientation-strain map using the corrected data, and the new multi-channel map was input to step V resulting in a final reconstructed single-channel orientation-strain map.

\subsubsection*{Algorithm summary}
We indexed orientation-strain states on a voxel grid, allowing for multiple indexed candidates per voxel. Subsequently, we refined the indexed multi-channel orientation-strain map based on voxel positions and used a median filtering operation to select an approximate solution. Using the approximate solution, we corrected our diffraction data and reran our refinement analysis and filtering algorithm. The final result is a voxelated single-channel orientation-strain map.

Our approach ensures that the final reconstruction is among the set of possible indexed solutions, adhering to diffraction data, while simultaneously suppressing noise by leveraging  the assumption of spatial correlation among local crystal orientations. It's worth noting that the median filtering operations are conducted with a filter size much smaller than the typical grain size, ensuring the ability to resolve spatial variations across the grains. However, strain and orientation variations localised on an even smaller scale, such as in the vicinity of boundaries, may be subject to larger errors, as discussed in section \ref{sec-stress_resolution}. 

It is also worth to mention that since the sample cross-section is relatively large compared with the detector-to-sample distance (0.24mm respectively 98.9mm), our correction for the spatial origin of diffraction not only suppresses spurious strain gradients that arise due to the finite sample size but also help to select the correct grain orientation locally during indexing (step III).

\subsection{Stress conversion}
\label{sec-stress_conversion}

We converted the strain tensor fields to stress tensor fields, voxel by voxel, using the elastic constants of Aluminum, \(D_{11}=104\) GPa, \(D_{12}=73\) GPa, and \(D_{44}=32\) GPa \cite{deJong2015}. Since the material parameters are given in relation to the IEEE standard \cite{IEEE1988}, conversion from strain to stress must take place in a local crystal coordinate system that aligns the Cartesian axes with the cubic unit cell axes. Explicitly, the fourth order elasticity tensor, given in the crystal reference frame, was taken as
\begin{equation}
    D = \begin{bmatrix}
        104 & 73 & 73 & 0 & 0 & 0 \\
        73 & 104 & 73 & 0 & 0 & 0 \\
        73 & 73 & 104 & 0 & 0 & 0 \\
        0 & 0 & 0 & 32 & 0 & 0 \\
        0 & 0 & 0 & 0 & 32 & 0 \\
        0 & 0 & 0 & 0 & 0 & 32
    \end{bmatrix} \text{GPa}.
\end{equation}
The strain tensor was converted to crystal coordinates as
\begin{equation}
    \boldsymbol{\epsilon}^{(c)} = \boldsymbol{U}^T \boldsymbol{\epsilon} \boldsymbol{U},
\end{equation}
using the local orientation matrix, \(\boldsymbol{U}\), associated to the voxel. The corresponding crystal coordinate stress components in the considered voxel could then be computed from Hooke's law as
\begin{equation}
    \begin{bmatrix}
        \sigma^{(c)}_{xx} \\
        \sigma^{(c)}_{yy} \\
        \sigma^{(c)}_{zz} \\
        \sigma^{(c)}_{yz} \\
        \sigma^{(c)}_{xz} \\
        \sigma^{(c)}_{xy} 
    \end{bmatrix} = 
    \begin{bmatrix}
        104 & 73 & 73 & 0 & 0 & 0 \\
        73 & 104 & 73 & 0 & 0 & 0 \\
        73 & 73 & 104 & 0 & 0 & 0 \\
        0 & 0 & 0 & 32 & 0 & 0 \\
        0 & 0 & 0 & 0 & 32 & 0 \\
        0 & 0 & 0 & 0 & 0 & 32
    \end{bmatrix}
    \begin{bmatrix}
        \epsilon^{(c)}_{xx} \\
        \epsilon^{(c)}_{yy} \\
        \epsilon^{(c)}_{zz} \\
        2\epsilon^{(c)}_{yz} \\
        2\epsilon^{(c)}_{xz} \\
        2\epsilon^{(c)}_{xy} \\
    \end{bmatrix}\text{GPa},
\end{equation}
where superscript \((c)\) denotes crystal reference frame. Conversion back to sample coordinates was then performed as
\begin{equation}
    \boldsymbol{\sigma} = \boldsymbol{U} \boldsymbol{\sigma}^{(c)} \boldsymbol{U}^T.
\end{equation}
Note that for materials that do not exhibit the cubic symmetry the above conversion between sample and crystal frame should feature an additional matrix transformation to ensure consistency with the IEEE conventions.

For completeness, we provide explicitly the definitions for the equivalent tensile stress 
\begin{equation}
    \sigma_e = \sqrt{3 J_2},
\end{equation}
where the second stress invariant, \(J_2\), is defined as
\begin{equation}
    J_2 = \sqrt{\dfrac{1}{2}\sum_{i}\sum_{j} s_{ij}s_{ji}},
\end{equation}
and the deviatoric stress, \(s_{ij}\), is given by
\begin{equation}
    s_{ij} = \sigma_{ij} - \dfrac{1}{3}\sigma_m\delta_{ij},
\end{equation}
where \(\delta_{ij}\) is Kronecker's delta, and the hydrostatic stress, \(\sigma_m\), is defined as
\begin{equation}
    \sigma_m = \dfrac{1}{3}(\sigma_{11} + \sigma_{22} + \sigma_{33}).
\end{equation}
Alphabetical and numerical subscripts for strain and stress tensors are equivalent, given by
\begin{equation}
    \boldsymbol{\sigma} = \begin{bmatrix}
        \sigma_{11} & \sigma_{12} & \sigma_{13} \\
        \sigma_{21} & \sigma_{22} & \sigma_{23} \\
        \sigma_{31} & \sigma_{32} & \sigma_{33} \\
    \end{bmatrix} = 
\begin{bmatrix}
        \sigma_{xx} & \sigma_{xy} & \sigma_{xz} \\
        \sigma_{yx} & \sigma_{yy} & \sigma_{yz} \\
        \sigma_{zx} & \sigma_{zy} & \sigma_{zz} \\
    \end{bmatrix}.
\end{equation}

\subsection{KAM filter \& Grain Boundary Identification}
\label{sec-KAM_filter}

\begin{figure}[H]
    \centering \includegraphics{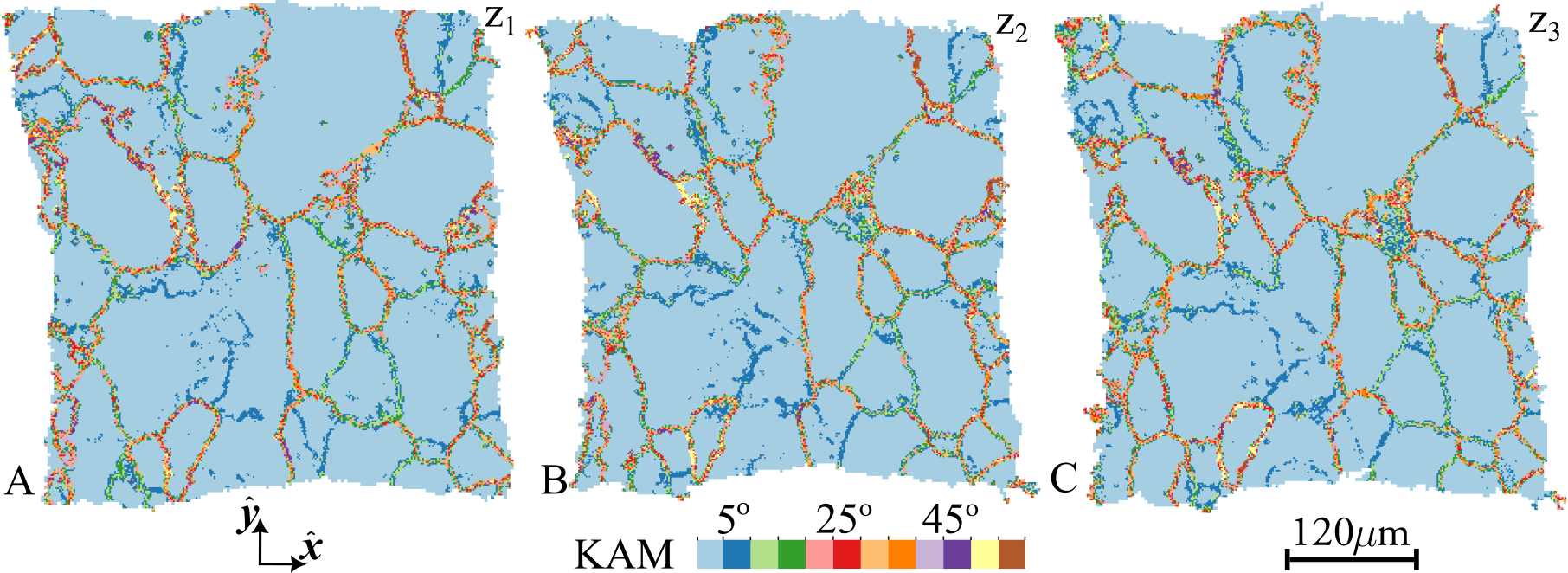}
    \caption{\textbf{KAM maps} for the three reconstructed \(z\)-layers, z1,z2 and z3 (A,B,C, respectively). }
    \label{fig:kam}
\end{figure}

\begin{figure}
    \centering \includegraphics{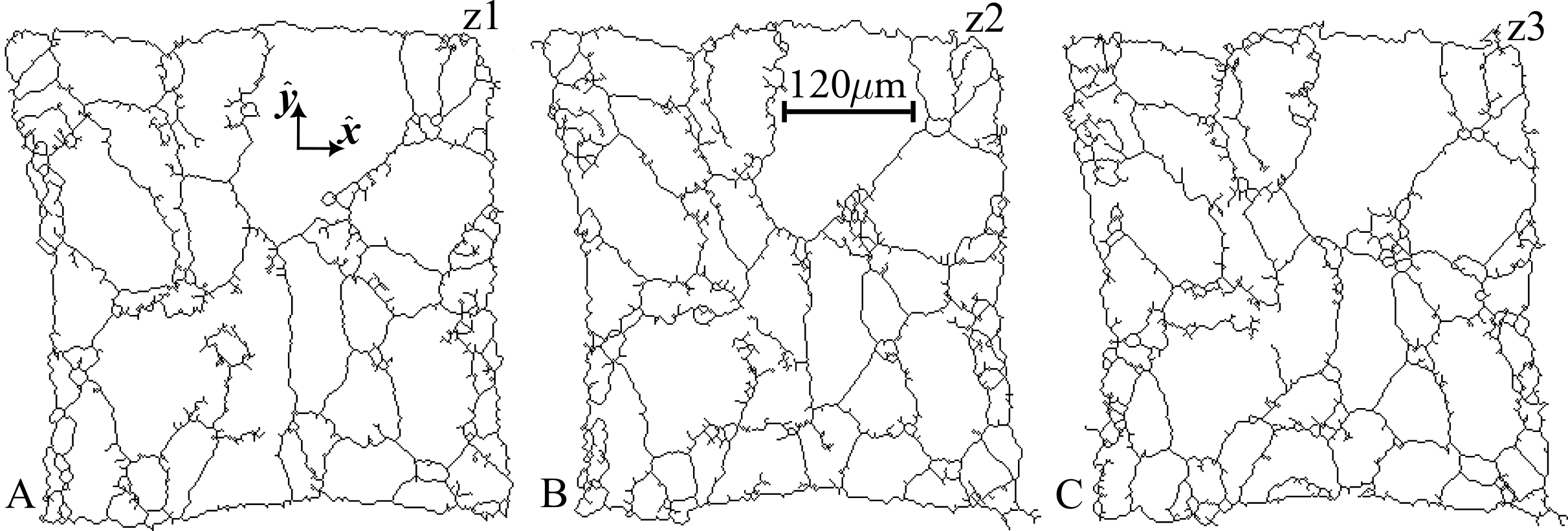}
    \caption{\textbf{Grain boundary skeletons} obtained from the KAM maps in fig. \ref{fig:kam}. }
    \label{fig:segmented_grain_boundaries}
\end{figure}

Within each slice in \(z\), Kernel average misorientations (KAM) \cite{Wright2011KAM} were computed, considering a neighborhood defined by the beam-size (3\(\mu\)m) and a lower misorientation threshold of \(1.8^o\) (fig. \ref{fig:kam}). Specifically, the neighborhood matrix was taken as
\begin{equation}
    \begin{bmatrix}
        0 & 0  &1 & 0 & 0 \\
        0 & 1 & 1 & 1 & 0 \\ 
        1 & 1 & 1 & 1 & 1 \\
        0 & 1 & 1 & 1 & 0 \\ 
        0 & 0 & 1 & 0 & 0 
        \end{bmatrix}.
\end{equation}
The resulting KAM maps are shown in fig. \ref{fig:kam}. Grain and sub-grain boundaries were identified for each \(z\)-layer separately by:
\begin{enumerate}
    \item Binarising the KAM map at a threshold of 4\(^o\).
    \item Applying a morphological thinning to the binary map.
    \item  Pruning the binary skeleton of small residual objects (min feature size 64 pixels with a 2-pixel connectivity).
\end{enumerate}
The resulting binary grain boundary skeletons are presented in fig. \ref{fig:segmented_grain_boundaries}.

\subsection{Grain orientations and intra-grain orientation distributions}
\label{sec_grain_orientations}
\begin{figure}[H]
    \centering
    \includegraphics[scale=1.0]{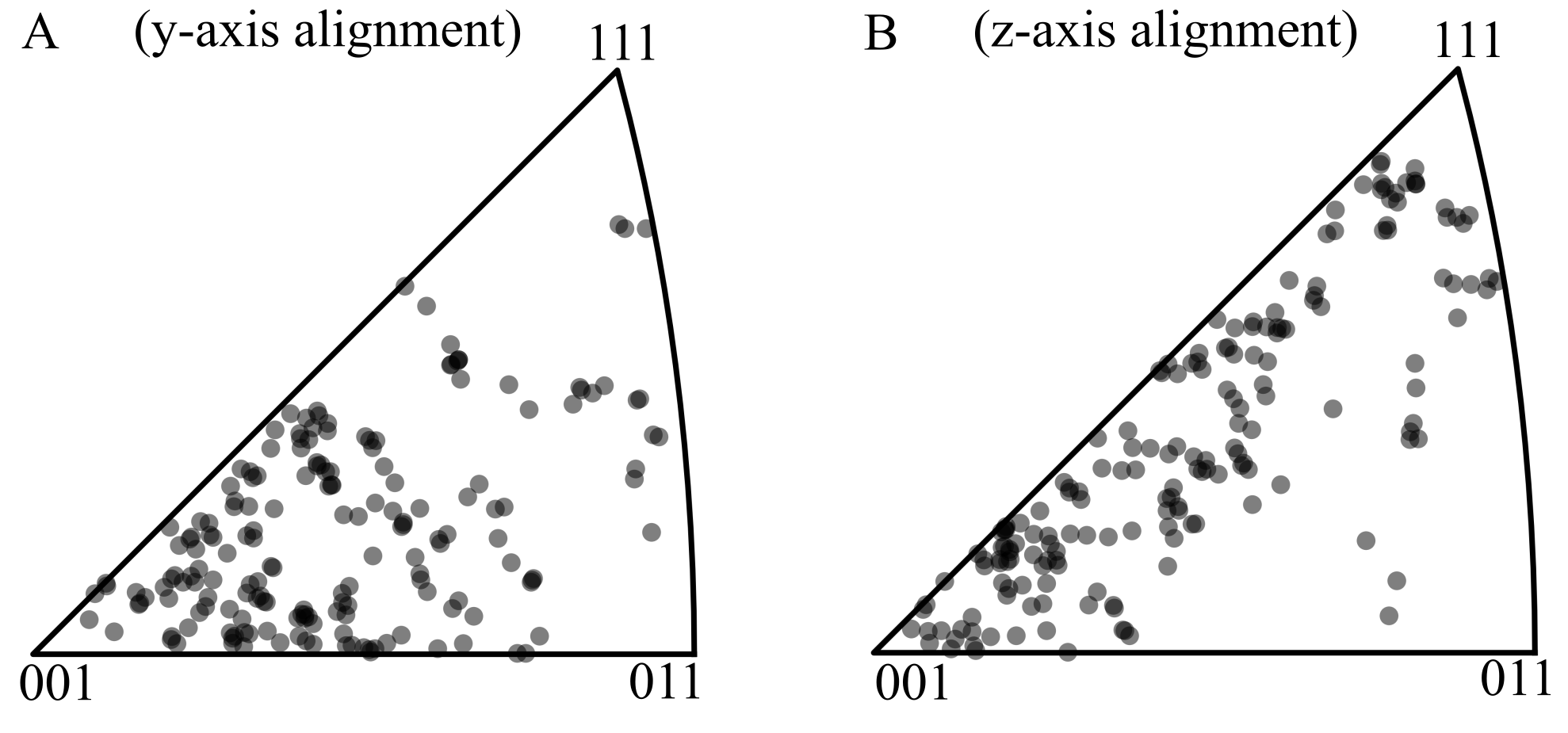}
    \caption{\textbf{Macroscopic texture}. Inverse pole figures of grain mean orientations with respect to (A) \(y\)-axis alignment and (B) \(z\)-axis alignment. The lattice plane normal to the 011 family is seen to rarely align with the \(z\)-axis (which is also the tensile axis), as is typical after tensile deformation. Each point in the inverse pole figure corresponds to the mean orientation of a segmented \(x\)-\(y\) grain-slice.}
    \label{fig:macro_ipf}
\end{figure}

 For each \(z\)-layer, all connected component regions found from the analysis in section \ref{sec-KAM_filter}, containing more than 25 voxels,  were identified as grains, and the mean unit cell over each such grain was computed. The resulting average orientations (each associated to one \(z\)-slice of one segmented grain) are plotted as inverse pole figures in fig. \ref{fig:macro_ipf}. The total number of represented grain \(z\)-slices is 179. As seen in fig. \ref{fig:macro_ipf}, this procedure revealed a typical tensile texture where the crystallographic direction aligned with the tensile axis (\(z\)-axis) spans the (001)-(111) line of the triangle for most grains, and grains with tensile direction along (011) are scarce.

\begin{figure}[H]
    \centering
    \includegraphics[scale=1.0]{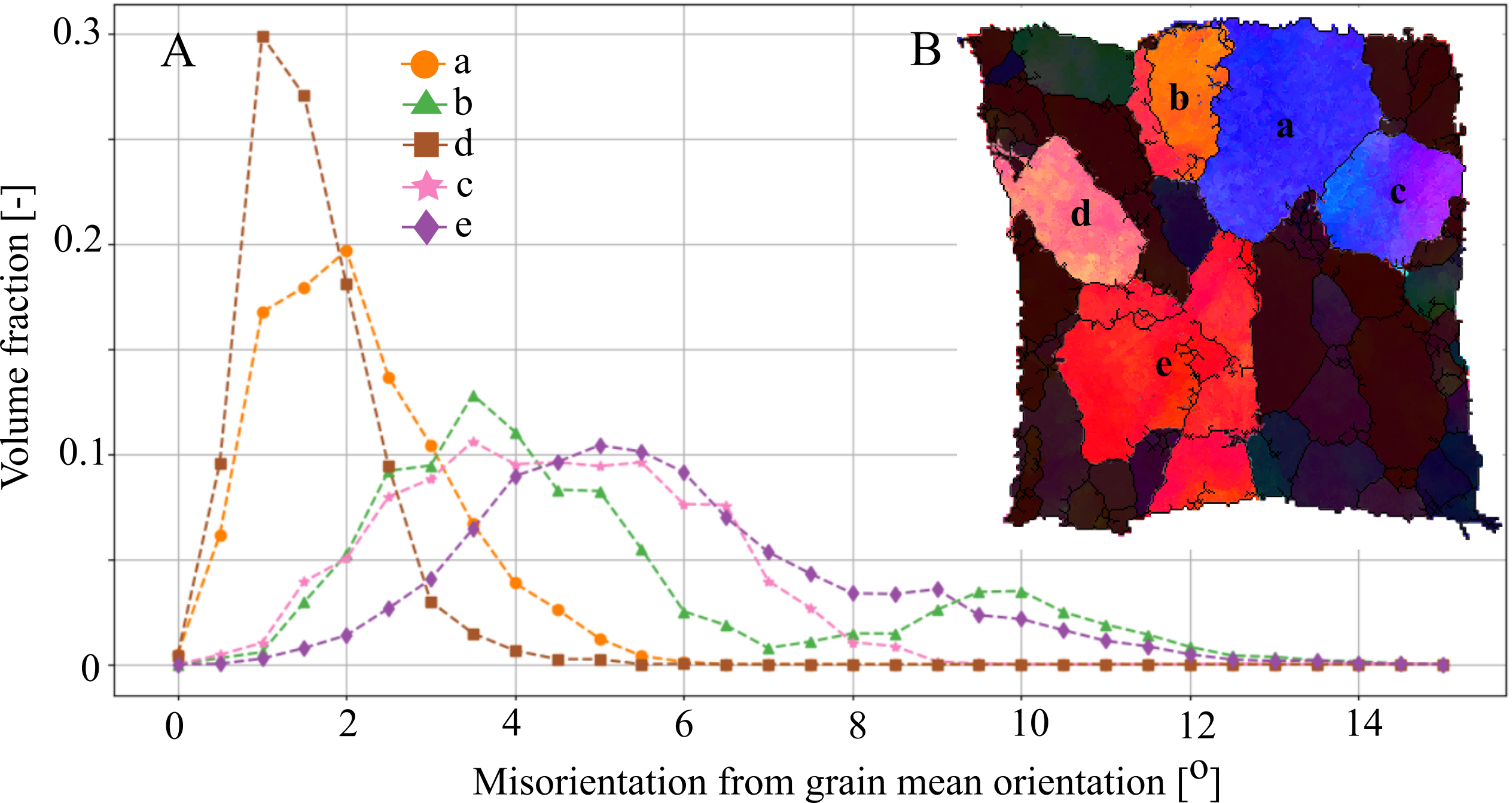}
    \caption{\textbf{Intra-grain misorientations}. Histograms of misorientation (A) for the five largest grains in the central layer (z2) of the sample (B, a-e). Each histogram (a-e) shows misorientations in relation to the grain mean orientation. Grain volumes were segmented using a flood fill approach, searching for spatially connected regions with local misorientations of less than 4.0\(^\circ\).}
    \label{fig:misori}
\end{figure}

The Intra-grain misorientation distributions around the grain average orientation showed significant broadening, as shown in fig. \ref{fig:misori}. Particularly, elevated levels of misorientation above 10~\(^o\) were observed in two large grains that featured multiple low-angle sub-grain boundaries (fig. \ref{fig:misori} grains b and e). The presence of orientation gradients in the orientation map are indicators of plastic strain, characterised by slipping of crystal planes, causing the lattice to warp, which subsequently resulted in severe diffraction peak arcing and diffraction peak overlap.

\subsection{Strain Tensor Maps}
\label{sec-strain_tensor_maps}

\begin{figure}[H]
    \centering
    \includegraphics[scale=0.95]{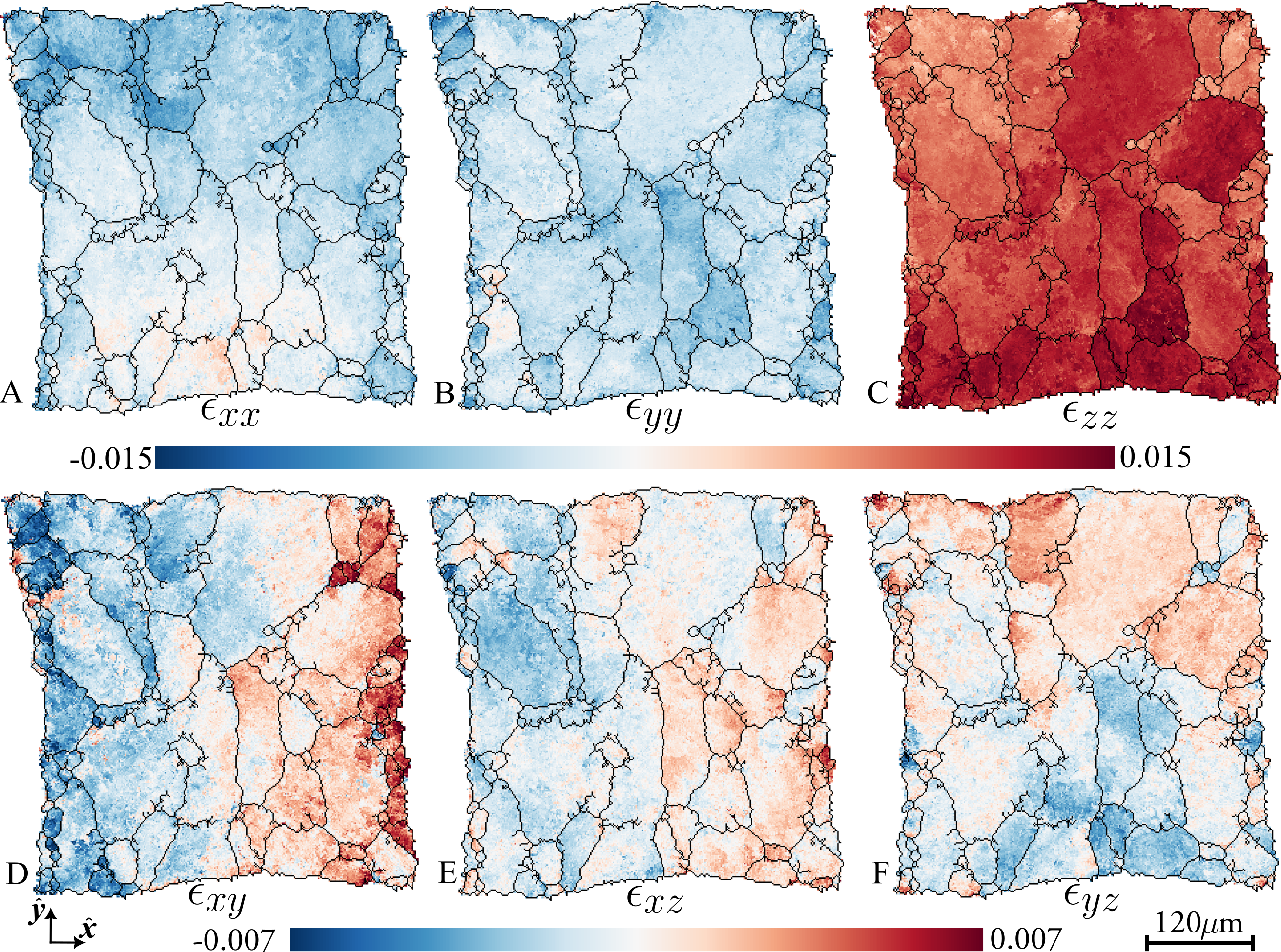}
    \caption{\textbf{Reconstructed strain tensor field in bottom (z1) layer}. Axial strain, \(\epsilon_{xx}\), \(\epsilon_{yy}\), \(\epsilon_{zz}\) (A-C), and shear strain, \(\epsilon_{xy}\), \(\epsilon_{xz}\), \(\epsilon_{yz}\) (D-F).}
    \label{fig:strain_bottom_layer}
\end{figure}

\begin{figure}[H]
    \centering
    \includegraphics[scale=0.95]{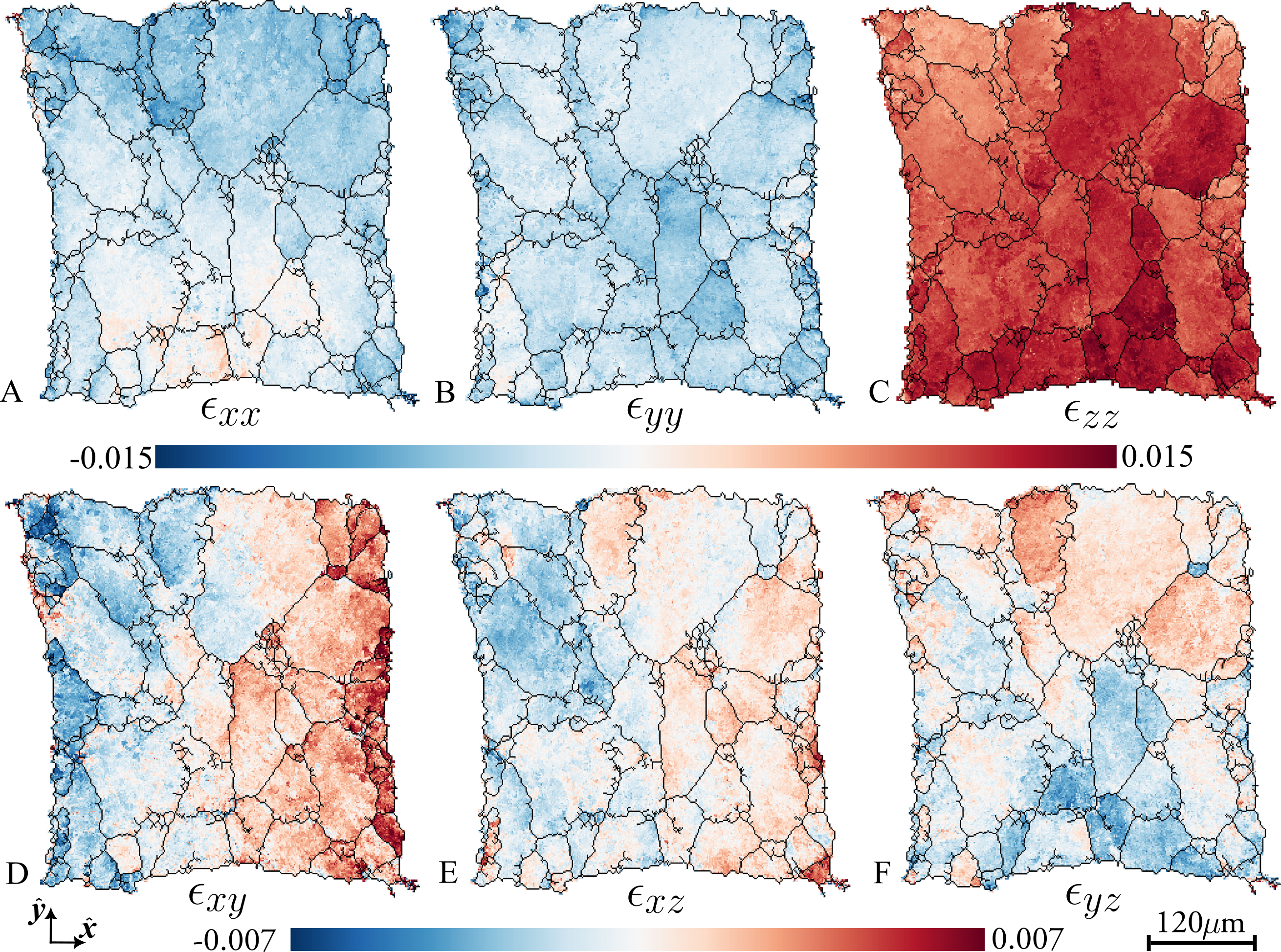}
    \caption{\textbf{Reconstructed strain tensor field in central (z2) layer}. Axial strain, \(\epsilon_{xx}\), \(\epsilon_{yy}\), \(\epsilon_{zz}\) (A-C), and shear strain, \(\epsilon_{xy}\), \(\epsilon_{xz}\), \(\epsilon_{yz}\) (D-F).}
    \label{fig:strain_center_layer}
\end{figure}

\begin{figure}[H]
    \centering
    \includegraphics[scale=0.95]{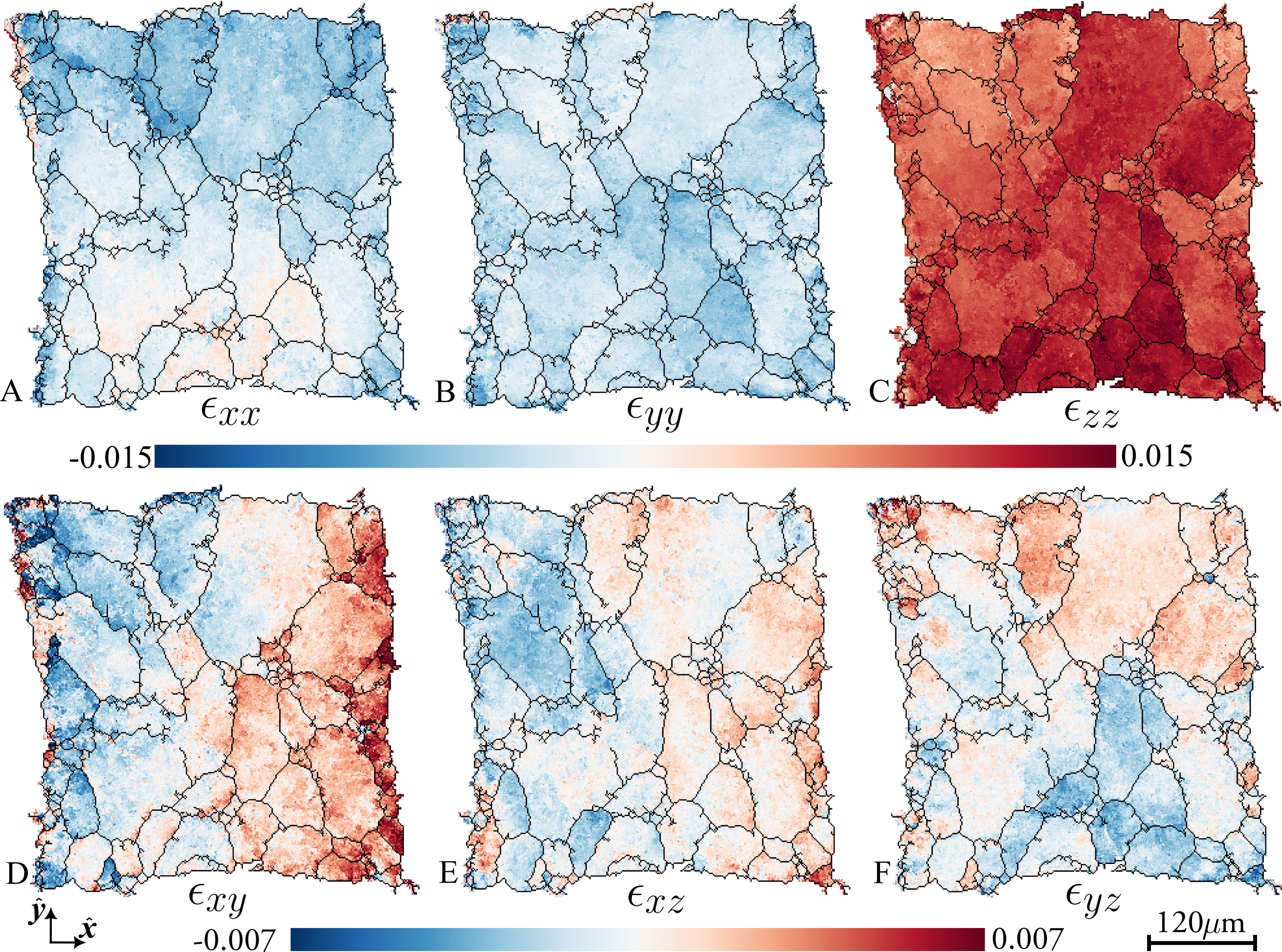}
    \caption{\textbf{Reconstructed strain tensor field in top (z3) layer}. Axial strain, \(\epsilon_{xx}\), \(\epsilon_{yy}\), \(\epsilon_{zz}\) (A-C), and shear strain, \(\epsilon_{xy}\), \(\epsilon_{xz}\), \(\epsilon_{yz}\) (D-F).}
    \label{fig:strain_top_layer}
\end{figure}

The full strain tensor field, reconstructed using the algorithm described in section \ref{sec-orientation-strain-reconstruction}, is presented in figs. \ref{fig:strain_bottom_layer}-\ref{fig:strain_top_layer}. The three figures illustrates the three separate layers, z1, z2 and z3. 

\newpage
\subsection{Stress Tensor Maps}
\label{sec-stress_tensor_maps}

\begin{figure}[H]
    \centering
    \includegraphics[scale=0.95]{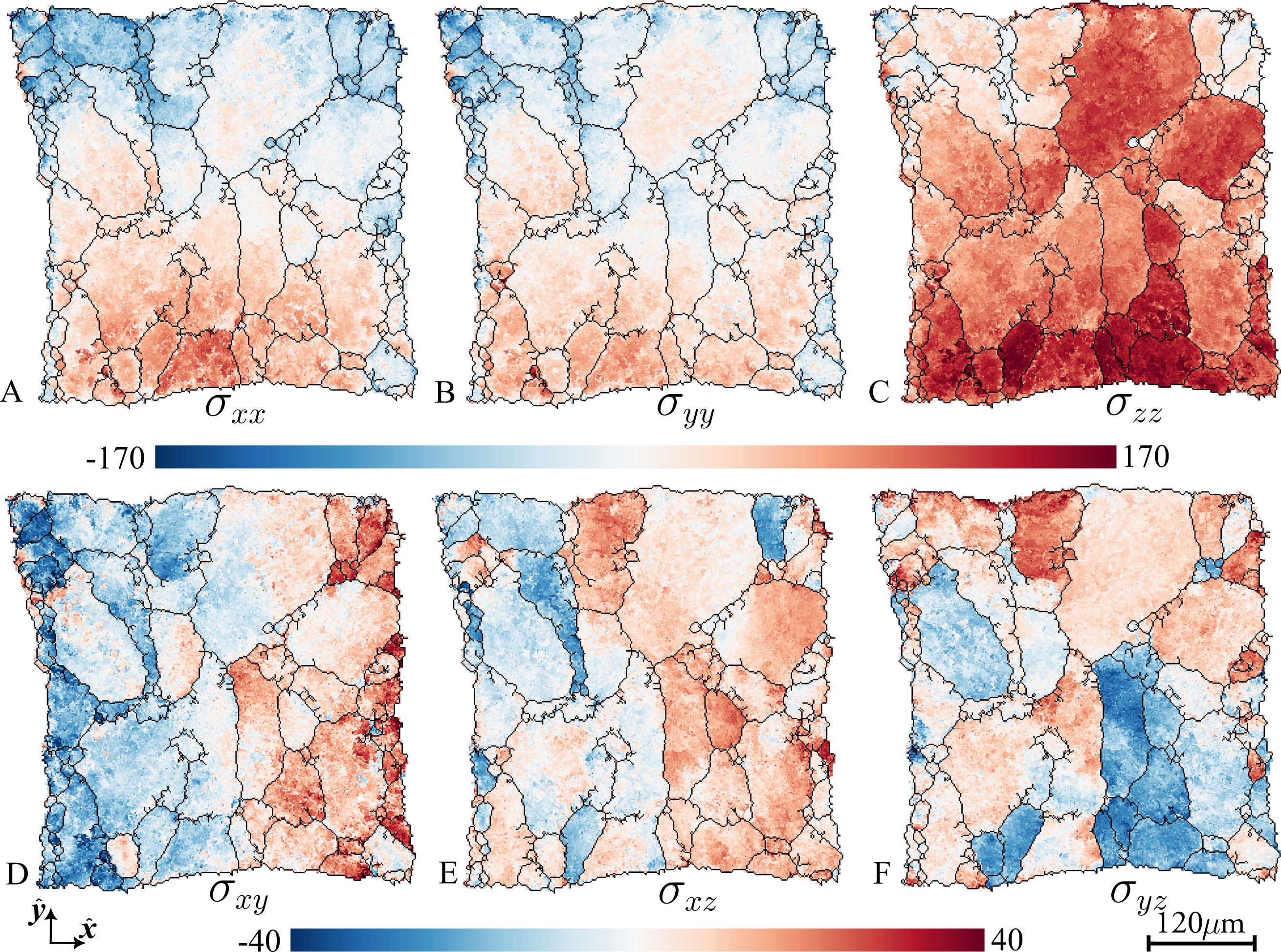}
    \caption{\textbf{Reconstructed stress tensor field in bottom (z1) layer}. Axial stress, \(\sigma_{xx}\), \(\sigma_{yy}\), \(\sigma_{zz}\) (A-C), and shear stress, \(\sigma_{xy}\), \(\sigma_{xz}\), \(\sigma_{yz}\) (D-F), stress in the bottom (z1) layer.}
    \label{fig:stress_bottom_layer}
\end{figure}

\begin{figure}[H]
    \centering
    \includegraphics[scale=0.95]{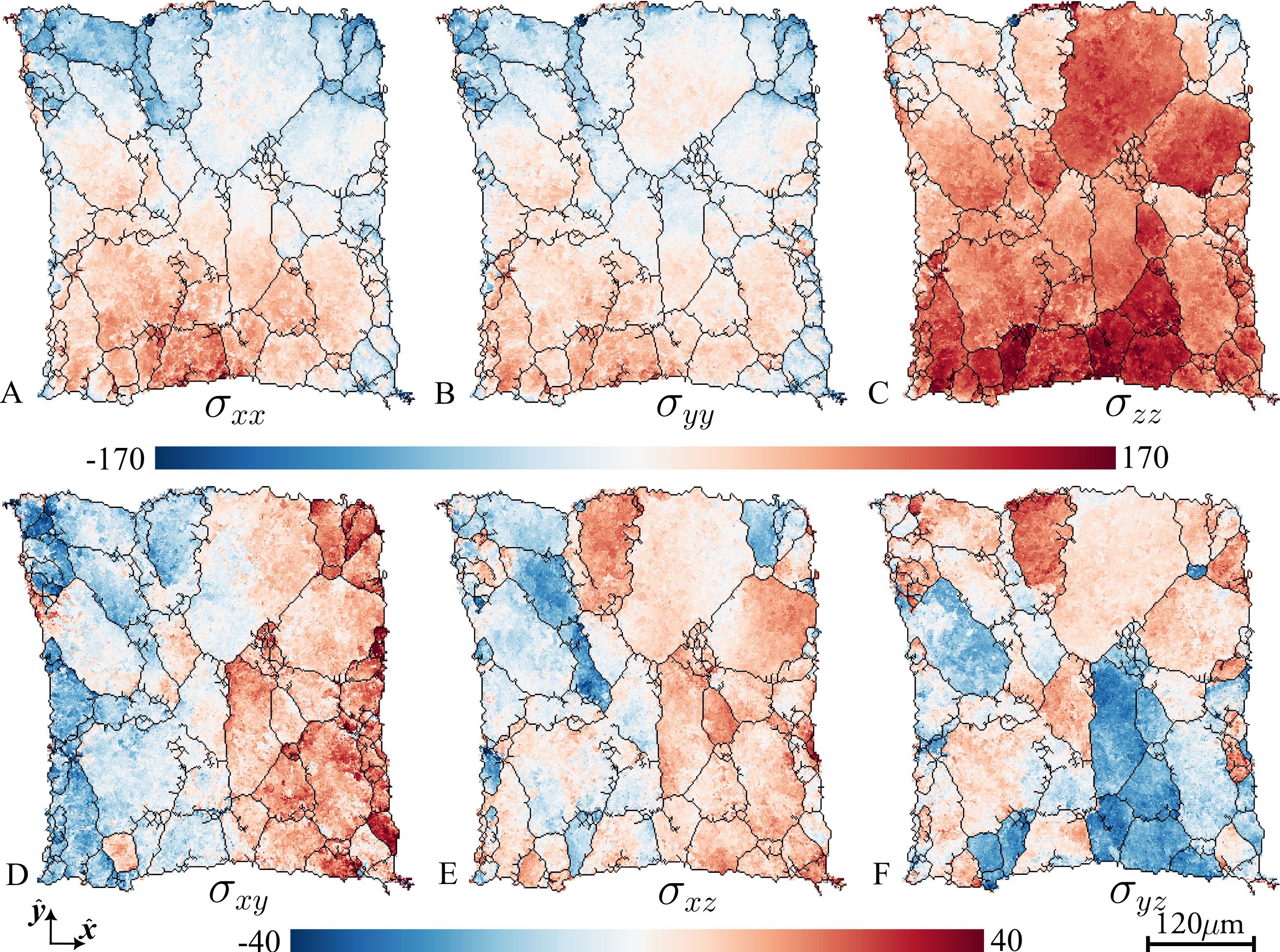}
    \caption{\textbf{Reconstructed stress tensor field in central (z2) layer}. Axial stress, \(\sigma_{xx}\), \(\sigma_{yy}\), \(\sigma_{zz}\) (A-C), and shear stress, \(\sigma_{xy}\), \(\sigma_{xz}\), \(\sigma_{yz}\) (D-F). This figure reproduces parts of the results in fig. 4 of the main paper. }
    \label{fig:stress_center_layer}
\end{figure}

\begin{figure}[H]
    \centering
    \includegraphics[scale=0.95]{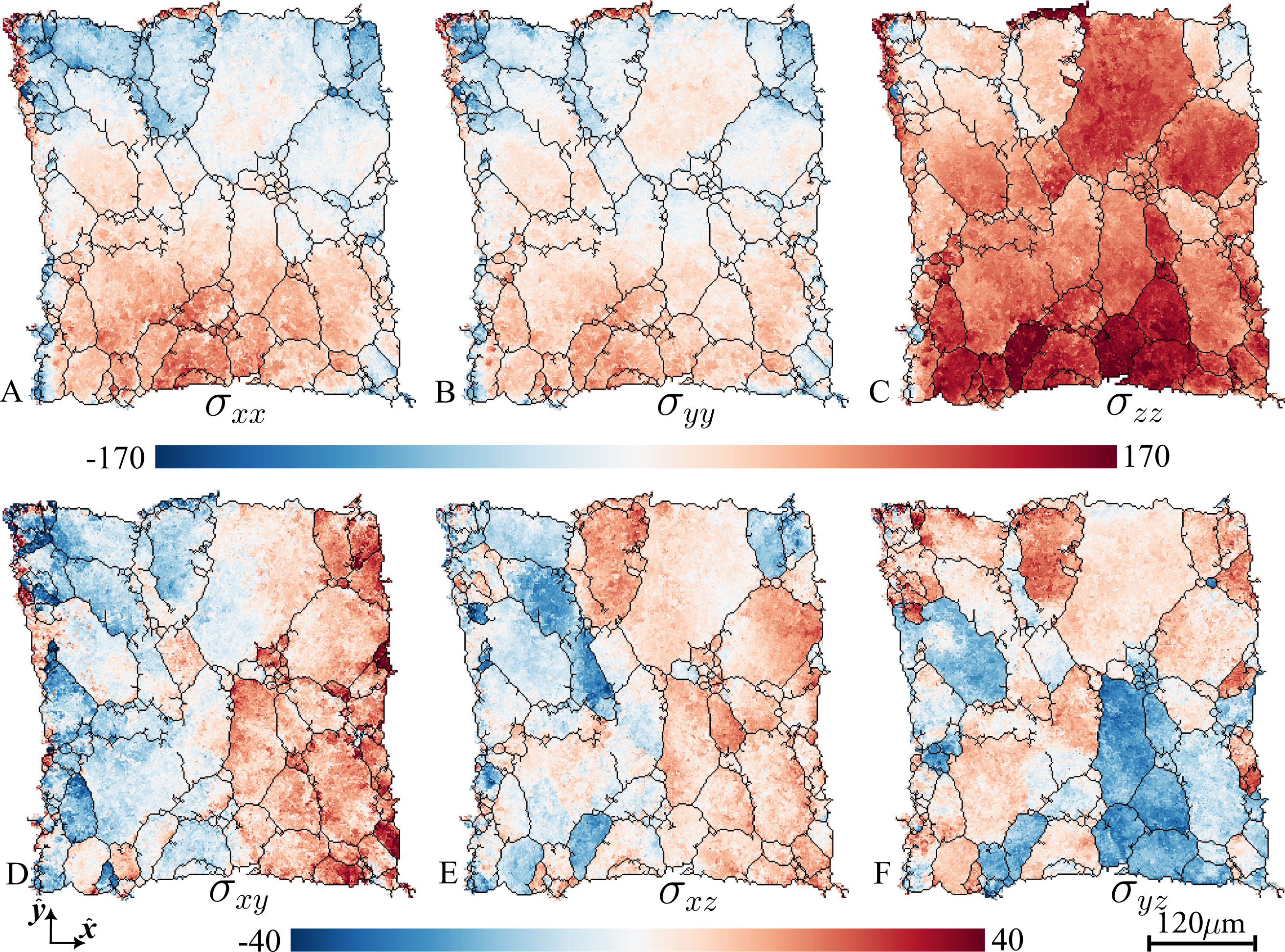}
    \caption{\textbf{Reconstructed stress tensor field in top (z3) layer}. Axial stress, \(\sigma_{xx}\), \(\sigma_{yy}\), \(\sigma_{zz}\) (A-C), and shear stress, \(\sigma_{xy}\), \(\sigma_{xz}\), \(\sigma_{yz}\) (D-F), .}
    \label{fig:stress_top_layer}
\end{figure}

The full stress tensor field, computed from the strain tensor fields in figs. \ref{fig:strain_bottom_layer}-\ref{fig:strain_top_layer}, are shown as figs. \ref{fig:stress_bottom_layer}-\ref{fig:stress_top_layer}. These complements the main material of the paper by showcasing the stress tensor for all three layers, z1, z2 and z3.

\newpage
\section{Error Estimation}
\label{sec-error_estimation}
\subsection{Spatial Resolution}
\label{sec-spatial_resolution}

\begin{figure}[H]
    \centering
    \includegraphics[scale=1.0]{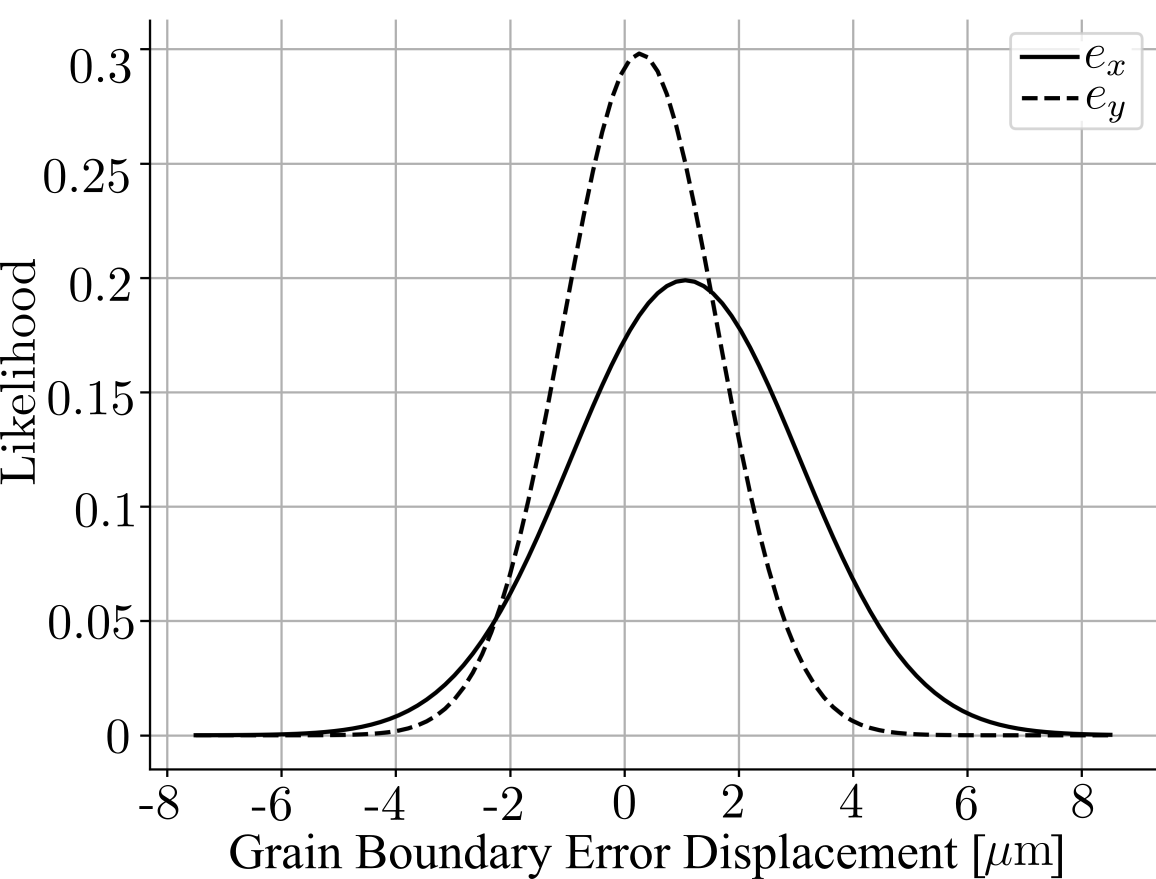}
    \caption{\textbf{Estimated spatial reconstruction error}. The likelihood of error displacements at grain boundaries in the reconstruction is shown. The error standard deviation is close to the voxel size ($\sim$1.5$\mu$m).}
    \label{fig:error}
\end{figure}

\begin{figure}[H]
    \centering
    \includegraphics[scale=1.0]{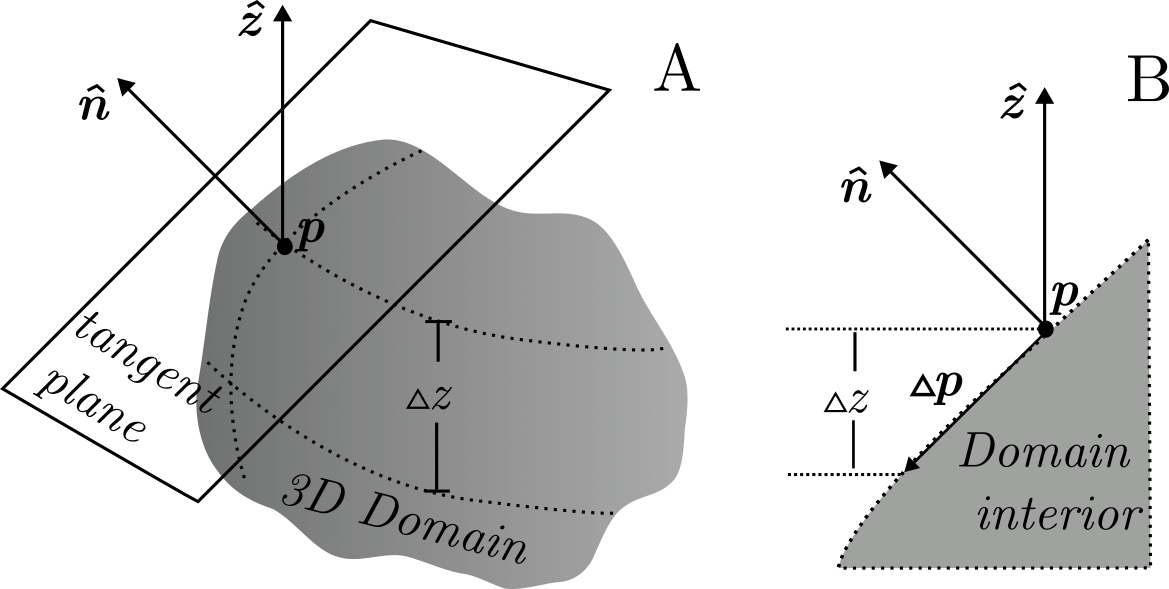}
    \caption{\textbf{Grain boundary curvature postulate}. A 3D domain has a tangent plane at \(\boldsymbol{p}\) with normal \(\boldsymbol{\hat{n}}\) (A). For an increment \(\Delta z \) along \(\boldsymbol{\hat{z}}\) the closest point to \(\boldsymbol{p}\) on the domain boundary is \(\boldsymbol{p}+\Delta\boldsymbol{p}\) (B).}
    \label{fig:depict}
\end{figure}

To estimate the spatial resolution in our reconstruction, we utilised the displacements of grain boundaries between consecutive \(z\)-layers in the grain volume. These displacements arise from a combination of the actual change in grain shape and an unknown spatial reconstruction error. Assuming that these two effects are uncorrelated and that the distribution of error displacements at grain boundaries follow a multivariate Gaussian distribution, we could infer the spatial resolution by comparing the reconstructed grain boundary skeletons of consecutive \(z\)-layers. We found an error standard deviation close to the voxel size ($\sim$1.5$\mu$m). For further details and the statistical characteristics of the error, refer to table \ref{tab:error_stats}. Profiles illustrating the estimated grain boundary error displacement distribution are presented in fig. \ref{fig:error}. Below, we provide a comprehensive explanation of the methodology employed to obtain this estimate.
\begin{table}[htbp]
    \centering
    \begin{tabular}{cc}
    \textbf{Metric} & \textbf{Value} \\
    \hline
        Standard deviation in x & $2.01$ $\mu$m \\
        \hline
        Standard deviation in y & $1.34$ $\mu$m \\
        \hline \\
        Mean & $\begin{bmatrix}1.06, 0.27\end{bmatrix}^T$ $\mu$m\\  \\
        \hline \\
        Covariance & $\begin{bmatrix} 4.02 & 0.57 \\ 0.57 & 1.79  \end{bmatrix}$ $\mu$m$^2$ \\  \\
        \hline
    \end{tabular}
    \caption{Statistics of estimated grain boundary displacement error distribution.}
    \label{tab:error_stats}
\end{table}
 Assume \(\boldsymbol{p}\) denotes a point on a grain boundary and let \(\boldsymbol{\hat{n}}\) be the associated normal to the grain boundary tangent plane at \(\boldsymbol{p}\). Given an increment, \(\Delta z\), in the direction of \(\boldsymbol{\hat{z}}\), a point \(\boldsymbol{p}+\Delta \boldsymbol{p}\), is reached by following the tangent plane with normal \(\boldsymbol{\hat{n}}\) starting from \(\boldsymbol{p}\). The neighbouring grain boundary point \(\boldsymbol{p}+\Delta \boldsymbol{p}\) is here defined as the closest point (in a Euclidean sense) to \(\boldsymbol{p}\) at \(z=p_z + \Delta z\). The situation is geometrically depicted in fig. \ref{fig:depict}. 
Projecting \(\boldsymbol{\hat{z}}\) unto the tangent plane with normal \(\boldsymbol{\hat{n}}\) we find that
\begin{equation}
    \Delta \boldsymbol{p} = \bigg( \dfrac{ \Delta z }{1 - n^2_z}\bigg)\begin{bmatrix}-n_xn_z \\ -n_yn_z\\ 1 - n^2_z\end{bmatrix}.
\end{equation}
Clearly, the tangent plane normal, \(\boldsymbol{\hat{n}}\), must feature a minimal angle, \(\varphi\), to \(\boldsymbol{\hat{z}}\), such that over the increment \(\Delta z\) the distance traversed by \(\Delta \boldsymbol{p}\) is less than or equal to the largest observed grain boundary displacement. Beyond this constraint, without any additional prior information on the grain shape morphology, it is reasonable to assume that the distribution of tangent plane normals, \(\boldsymbol{\hat{n}}\), spans the allowable part of the unit ball uniformly. Specifically, we introduce the feasible set of tangent normals, \(\boldsymbol{\hat{n}} \in \mathcal{W}_{\phi}\), defined as
 \begin{equation}
     \boldsymbol{\hat{n}} \in \mathcal{W}_{\varphi} \quad \text{if} \quad \boldsymbol{\hat{n}} \in \mathcal{S}^3 \quad \text{and} \quad \arccos( |\boldsymbol{\hat{n}}^T\boldsymbol{\hat{z}}| ) > \varphi
 \end{equation}
 where \(\mathcal{S}^3\) is the shell of the unit ball in \(\mathbb{R}^3\). Our postulate then takes the form
\begin{equation}
    \boldsymbol{\hat{n}} \sim U(\mathcal{W}_{\chi})
\end{equation}
where \(U\) is the uniform distribution.

\begin{figure}[H]
    \centering
    \includegraphics[scale=0.9]{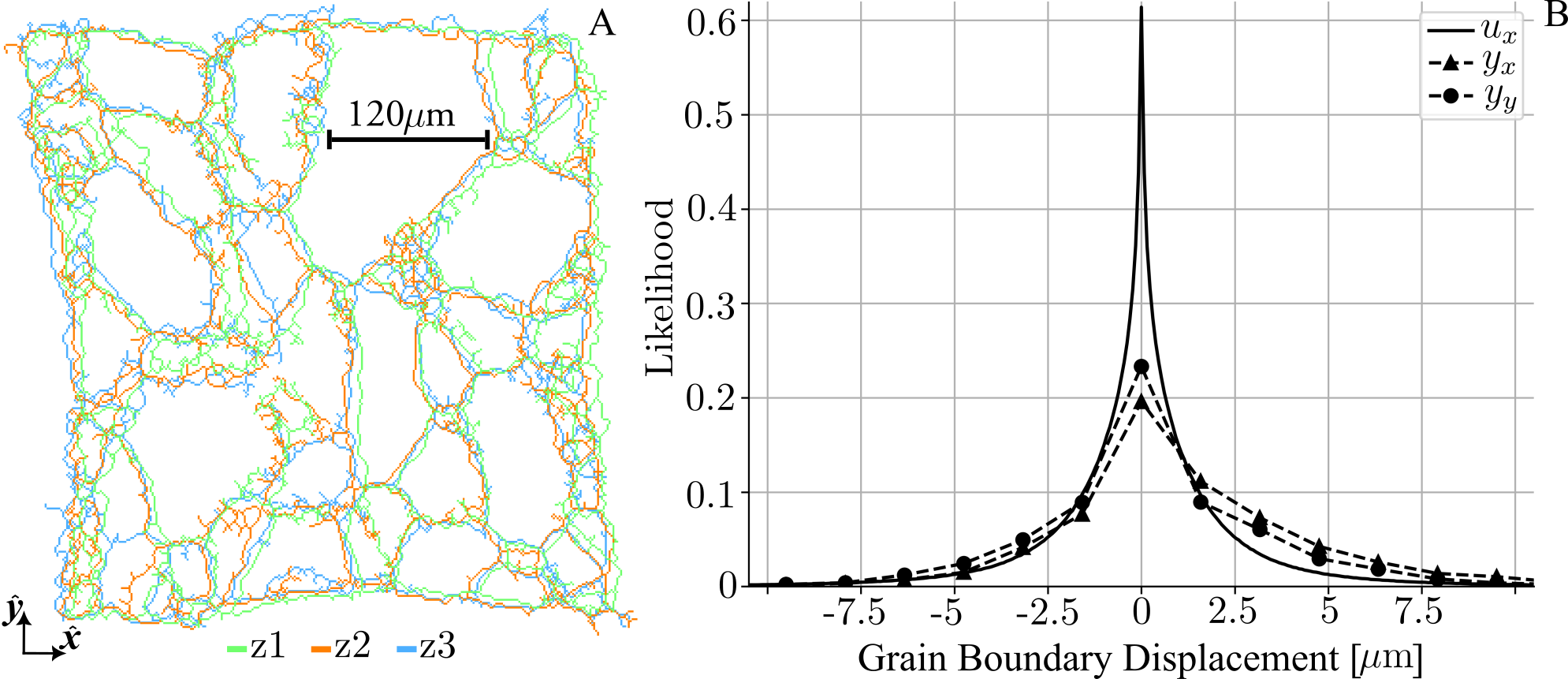}
    \caption{\textbf{Grain boundary displacement distributions}. Grain boundary skeletons obtained by morphological thinning of respective KAM-maps for the three reconstructed \(z\)-layers have been overlaid in green, orange and blue (A). Distributions over the measured grain boundary displacements, \(y_x, y_y\), are shown in (B) as dashed lines. Due to spatial errors in the reconstruction the distributions of  \(y_x\) and \(y_y\) show broadening compared to the expected set of displacements \(u_x\). This feature was exploited to estimate the spatial error of the reconstruction.}
    \label{fig:GBerror}
\end{figure}

Given that the tangent plane normal, \(\boldsymbol{\hat{n}}\), is a stochastic variable it follows that the vector \(\Delta \boldsymbol{p}\) is also a stochastic variable. We can therefore define a \(x\)-\(y\) planar stochastic displacement vector, \(\boldsymbol{u}\in\mathbb{R}^2\), as
\begin{equation}
    \boldsymbol{u} = \bigg( \dfrac{ \Delta z }{1 - n^2_z}\bigg)\begin{bmatrix}-n_xn_z \\ -n_yn_z\end{bmatrix}.
    \label{eq:u}
\end{equation}
The probability density function (PDF) of \(\boldsymbol{u}\) describes a distribution of displacements that originate from the postulated grain boundary curvature. We note that \(\boldsymbol{u}\) is isotropic in \(u_x\) and \(u_y\) with mean \(\boldsymbol{0}\). In fig. \ref{fig:GBerror} B (solid line) the PDF (profile) of \(\boldsymbol{u}\) is graphed by drawing 10 000 000 random samples from \eqref{eq:u}.

Let us now consider the case where an orientation-strain map has been reconstructed together with an accompanying grain boundary map. We defined a measured grain boundary displacement, \(\boldsymbol{y}\in\mathbb{R}^{2}\), by:
\begin{enumerate}
    \item Selecting a grain boundary point, \(\boldsymbol{p}\), in one of the reconstructed \(z\)-slices.
    \item Computing the candidate displacement vectors between \(\boldsymbol{p}\) and all grain boundary points at \(z + \Delta z\) (i.e in the consecutive \(z\)-slice).
    \item Selecting \(\boldsymbol{y}\) as the displacement vector with the minimal euclidean norm.
\end{enumerate}
Thus, the measurement, \(\boldsymbol{y}\), represents a noisy measurement of the true displacement, \(\boldsymbol{u}\). We used an additive error model
\begin{equation}
    \boldsymbol{y} = \boldsymbol{u} + \boldsymbol{e},
    \label{eq:y}
\end{equation}
where \(\boldsymbol{e}\in\mathbb{R}^2\) is the sought displacement error variable. When \(\boldsymbol{e}\rightarrow 0\) we expect \(\boldsymbol{y}\) to be a draw representative of \(\boldsymbol{u}\), and when \(\boldsymbol{e}\) is large we expect the distribution of \(\boldsymbol{y}\) to depart from that of \(\boldsymbol{u}\). The grain boundary displacement error, \(\boldsymbol{e}\), is here postulated as an aggregated displacement error that stems from, possibly, multiple sources, such as finite data size, data noise, outliers, forward model inconsistency, loss of precision etc. Without any further knowledge we take \(\boldsymbol{e}\) to be multivariate Gaussian
\begin{equation}
    \boldsymbol{e} \sim \mathcal{N}(\boldsymbol{\mu}, \boldsymbol{\Sigma})
\end{equation}
Given a series of measurements \(\boldsymbol{y}_1,\boldsymbol{y}_2,...,\boldsymbol{y}_N\), the task is now to estimate the parameters, \(\boldsymbol{\mu},\boldsymbol{\sigma}\), of the error distribution. Letting \(\mathbb{E}[]\) be the mean operator we find that
\begin{equation}
    \mathbb{E}[\boldsymbol{e}] = \mathbb{E}[\boldsymbol{y}-\boldsymbol{u}] = \mathbb{E}[\boldsymbol{y}],
    \label{eq:mean}
\end{equation}
and Letting \(\mathbb{V}[]\) be the variance operator we find that
\begin{equation}
    \mathbb{V}[\boldsymbol{e},\boldsymbol{e}] = \mathbb{V}[\boldsymbol{y},\boldsymbol{y}]- \mathbb{V}[\boldsymbol{u},\boldsymbol{u}],
    \label{eq:var}
\end{equation}
where it was used that \(\boldsymbol{u}\) is zero mean. To proceed in estimating the error parameters, \(\boldsymbol{\mu},\boldsymbol{\sigma}\), we must first determine the unknown angle \(\varphi\) which defines the distribution of \(\boldsymbol{u}\). We approximated \(\varphi\) from displacement data, \(\boldsymbol{y}_1,\boldsymbol{y}_2,...,\boldsymbol{y}_N\), by taking the maximum displacement increment and computing the corresponding tangent plane normal angle needed to achieve this displacement,
\begin{equation}
    \varphi = \arctan\bigg( \dfrac{|\Delta z|}{\sqrt{\boldsymbol{y}^T_{max}\boldsymbol{y}_{max}}}\bigg ).
    \label{eq:phi}
\end{equation}
With \(\Delta z=\pm\)3\(\mu\)m equation \eqref{eq:phi} gave \(\varphi=\)7.77\(^o\).

With \(\varphi\) determined the quantities, \(\mathbb{V}[\boldsymbol{y},\boldsymbol{y}]\) and \(\mathbb{E}[\boldsymbol{y}]\) involved in equations \eqref{eq:mean} and \eqref{eq:var} were estimated from the displacement data using a maximum likelihood approach. Likewise, we sampled \(\boldsymbol{u}\) 10 000 000 times and estimated \(\mathbb{V}[\boldsymbol{u},\boldsymbol{u}]\) numerically using a maximum likelihood approach. The result of this procedure is found in table \ref{tab:error_stats}. In fig. \ref{fig:GBerror} the grain boundary skeletons are shown (A) together with the distribution of \(\boldsymbol{y}\) and \(\boldsymbol{u}\) in (B). We reiterate that, given our model assumptions, in an error free reconstruction, the distributions of \(\boldsymbol{y}\) and \(\boldsymbol{u}\) should be identical. The spatial errors manifest in fig. \ref{fig:GBerror} B as broadening in distributions of \(y_x\) and \(y_y\) compared to the postulated error free set of displacements \(u_x\).

\subsection{Stress Resolution}
\label{sec-stress_resolution}

\begin{figure}[H]
    \centering
    \includegraphics[scale=0.85]{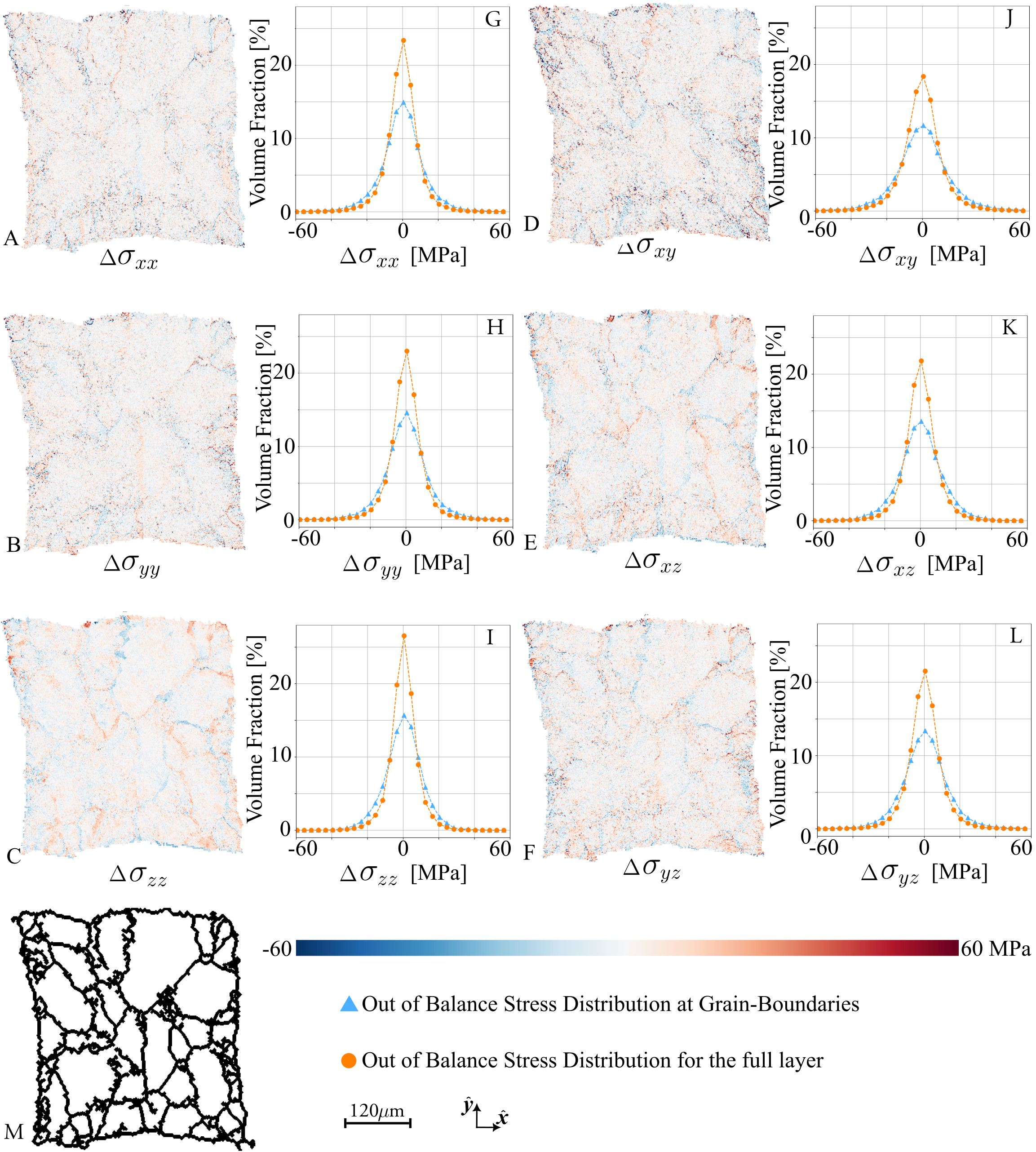}
    \caption{\textbf{Residual stress}. Out of balance stress tensor fields (A-F) and corresponding histograms (G-L) for the central slice (z2). Voxels close to grain-boundaries are marked in (M) and show a higher out of balance stress variance in the histograms (G-L). The out of balance stress is a measure of reconstruction error.}
    \label{fig:res_stress}
\end{figure}

Although we do not have access to a ground truth stress field, it is nevertheless possible to evaluate the accuracy of our reconstructed stress field by comparing to well established theory in the field of continuum mechanics. For a body in static equilibrium, in the absence of body forces, Cauchy's first law of motion demands that the divergence of stress vanish everywhere
\begin{equation}
    \text{div}(\sigma_{ij}) = \sum^{j=3}_{j=1} \dfrac{\partial \sigma_{ij}}{\partial x_j} = 0,
    \label{eq:divsig}
\end{equation}
where \(x_1=x\), \(x_2=y\) and \(x_3=z\). In our case we have access to a finite approximation of the continuous stress field with one stress tensor per voxel. To evaluate the differential terms in \eqref{eq:divsig} we must therefore use a finite difference approximation,
\begin{equation}
\begin{split}
    \dfrac{\partial \sigma_{ij}}{\partial x} \approx \dfrac{\sigma_{ij}(x+dx, y, z) - \sigma_{ij}(x, y, z)}{dx}, \\
    \dfrac{\partial \sigma_{ij}}{\partial y} \approx \dfrac{\sigma_{ij}(x, y+dy, z) - \sigma_{ij}(x, y, z)}{dy}, \\
    \dfrac{\partial \sigma_{ij}}{\partial z} \approx \dfrac{\sigma_{ij}(x, y, z+dz) - \sigma_{ij}(x, y, z)}{dz}. \\
    \label{eq:finitdiff}
\end{split}
\end{equation}
To make all voxel dimensions equal we interpolate our reconstructed volume linearly in \(z\) such that the finite increment \(dx=dy=dz=1.5\)\(\mu\)m can be used in \eqref{eq:finitdiff}. Inserting \eqref{eq:finitdiff} into \eqref{eq:divsig} and multiplying though with \(dx\) yields three (\(i=1,2,3\)) equations
\begin{equation}
\begin{split}
    r_i(\sigma_{ij}, x, y, z) = \sigma_{i1}(x+dx, y, z) - \sigma_{i1}(x, y, z) \\
    + \sigma_{i2}(x, y+dx, z) - \sigma_{i2}(x, y, z)  \\
    + \sigma_{i3}(x,y,z+dx) - \sigma_{i3}(x, y, z).
    \label{eq:r_i_expanded}
\end{split}
\end{equation}
When \(r_i(\sigma_{ij}, x, y, z)\neq0\) there exist an error in the stress field reconstruction and we seek to quantify how much the corresponding stress tensor, \(\sigma_{ij}(x,y,z)\), need to be perturbed to bring balance of forces. To this end we introduce the local stress increment, \(d\sigma_{ij}\), and define a perturbed stress
\begin{equation}
    \sigma'_{ij}  = \sigma_{ij}  + d\sigma_{ij}.
\end{equation}
From the linearity of \eqref{eq:r_i_expanded} it follows that any stress increment that satisfy
\begin{equation}
\begin{split}
    r_i(\sigma'_{ij}, x, y, z) = r_i(\sigma_{ij}, x, y, z) + r_i(d\sigma_{ij}, x, y, z) = \\
    r_i(\sigma_{ij}, x, y, z) - d\sigma_{i1} - d\sigma_{i2} - d\sigma_{i3} = 0,
\end{split}
\label{eq:balance}
\end{equation}
will bring balance of forces. To parameterise the solutions, \(d\sigma_{ij}\), to equation \eqref{eq:balance} we the introduce a column vector format
\begin{equation}
    d\boldsymbol{\Bar{\sigma}} = \begin{bmatrix}
        d\sigma_{11} \\
        d\sigma_{22} \\
        d\sigma_{33} \\
        d\sigma_{12} \\
        d\sigma_{13} \\
        d\sigma_{23}
    \end{bmatrix},\quad  \boldsymbol{r} =\begin{bmatrix}
        r_1(\sigma_{ij}, x, y, z)  \\
        r_2(\sigma_{ij}, x, y, z)  \\
        r_3(\sigma_{ij}, x, y, z)  
    \end{bmatrix}.
\end{equation}
Equation \eqref{eq:balance} can now be written as 
\begin{equation}
     \boldsymbol{r} = \boldsymbol{A}d\boldsymbol{\Bar{\sigma}}
     \label{eq:to_solve},
\end{equation}
where 
\begin{equation}
    \boldsymbol{A} = \begin{bmatrix}
        1 & 0 & 0 & 1 & 1 & 0 \\
        0 & 1 & 0 & 1 & 0 & 1 \\
        0 & 0 & 1 & 0 & 1 & 1 \\
    \end{bmatrix}.
\end{equation}
While \eqref{eq:to_solve} has many solutions, only one will uniquely minimise the Euclidean norm of the stress increment, \(d\boldsymbol{\Bar{\sigma}}\), namely the least squares solution
\begin{equation}
    d\boldsymbol{\Bar{\sigma}} = (\boldsymbol{A}^T\boldsymbol{A})^{-1}\boldsymbol{A}^T\boldsymbol{r}.
\end{equation}
By computing \(\boldsymbol{r}\) and solving for \(d\boldsymbol{\Bar{\sigma}}\) at every voxel in the reconstructed stress volume independently, an out of balance stress tensor voxel volume, \(\Delta \boldsymbol{\sigma}(x,y,z)\), could be defined. The result of this computation is shown in figs. \ref{fig:res_stress}A-F for the central slice (z2). The corresponding histograms over out of balance stress are shown in figs. \ref{fig:res_stress}G-L together with histograms of the out of balance stress close to grain boundaries. As shown in Fig. M, we here defined proximity to grain-boundaries by twice dilating the binary grain-boundary skeleton. Voxels close to grain boundaries showed a larger variance compared to the variance of the total distribution. The standard deviation for the respective distributions of Fig. \ref{fig:res_stress} are given in table \ref{tab:res_stress_std}. Comparing to figs. \ref{fig:stress_bottom_layer}-\ref{fig:stress_top_layer}, we note that while the absolute standard deviation across stress components are similar (8-12 MPa) the relative error in stress is lower for the axial stress components (\(\sigma_{xx},\sigma_{yy},\sigma_{zz}\)) compared to the shear stress (\(\sigma_{xy}, \sigma_{xz}, \sigma_{xz}\)).

Close to grain boundaries, multiple distinct grains diffract simultaneously, maximising diffraction peak overlap and making orientation classification challenging. The elevated error in stress close to grain boundaries can therefore be explained, in part, by local orientation reconstruction errors that propagate into stress owing to the anisotropic stiffness model. Beyond this, we emphasises that the true stress fields may feature sharp interfaces close to and at grain boundaries with local stress gradients not fully characterised at a finite spatial resolution.

\begin{table}[htbp]
    \centering
    \caption{Standard Deviation of Stress Components (MPa)}
    \begin{tabular}{lcccccc}
        & $\Delta \sigma_{xx}$ & $\Delta \sigma_{yy}$ & $\Delta \sigma_{zz}$ & $\Delta \sigma_{xy}$ & $\Delta \sigma_{xz}$ & $\Delta \sigma_{yz}$ \\
        Whole Layer & 9.76 & 9.73 & 7.82 & 12.28 & 10.00 & 10.02 \\
        Grain Boundary & 13.55 & 13.71 & 10.74 & 16.99 & 13.99 & 14.01 \\
    \end{tabular}
    \label{tab:res_stress_std}
\end{table}

\end{document}